\documentclass[aps,prb,twocolumn]{revtex4-2}
\usepackage[utf8]{inputenc} 
\usepackage[T1]{fontenc} 
\usepackage[english]{babel}
\usepackage[autostyle=true]{csquotes} 
\usepackage[dvipsnames]{xcolor}
\usepackage{ragged2e}
\usepackage{graphicx}
\usepackage{subcaption}
\usepackage{array}
\usepackage{amsmath} 
\usepackage{amssymb} 
\usepackage{amsfonts} 
\usepackage{amsbsy}
\usepackage{mathrsfs} 
\usepackage{dsfont} 
\usepackage{dcolumn} 
\usepackage{upgreek} 
\usepackage{bm} 
\usepackage{bbm}
\usepackage{placeins} 
\usepackage{physics} 
\usepackage[compat=1.1.0]{tikz-feynman}
\usetikzlibrary{arrows.meta,calc,fit,backgrounds,positioning}
\usepackage{booktabs} 
\usepackage[colorlinks,linkcolor=blue,citecolor=blue,urlcolor=blue,hypertexnames=false]{hyperref}
\usepackage[most]{tcolorbox}
\usepackage{float}
\usepackage[sort&compress]{natbib}
\usepackage[dvips]{epsfig}
\usepackage{hhline}
\usepackage{multirow}
\usepackage[normalem]{ulem}
\usepackage{xspace}
\def\ai	        {{\em ab--initio}}

\renewcommand{\[}{\left[}
\renewcommand{\]}{\right]}
\renewcommand{\(}{\left(}
\renewcommand{\)}{\right)}
\def\nl         {\right.\\ \left.}
\def\la         {\left\langle}
\def\ra         {\right\rangle}

\def\rar        {\rightarrow}  

\def\Rar        {\Rightarrow}

\newcommand{\olrar}[1]{\overleftrightarrow{#1}}

\newcommand{\eq}[1]{\begin{align}#1\end{align}}
\newcommand{\ml}[1]{\begin{multline}#1\end{multline}}

\newcommand{\eqg}[1]{\begin{gather}#1\end{gather}}

\newcommand{\seq}[1]{\begin{subequations}#1\end{subequations}}

\newcommand{\seql}[2]{\begin{subequations}\label{#1}#2\end{subequations}}
\newcommand{\seqal}[2]{\begin{subequations}\label{#1}\begin{align}#2\end{align}\end{subequations}}
\newcommand{\mll}[2]{\begin{multline}\label{#1}#2\end{multline}}
\newcommand{\eql}[2]{\begin{align}\label{#1}#2\end{align}}
\newcommand{\eqgl}[2]{\seq{\label{#1}\begin{gather}#2\end{gather}}}

\newcommand{\eqsl}[2]{\begin{equation}\begin{split}\label{#1}#2\end{split}\end{equation}}
\newcommand{\stkout}[1]{\ifmmode\text{\sout{\ensuremath{#1}}}\else\sout{#1}\fi}
\newcommand{\average}[1]{\left\langle #1 \right\rangle}
\newcommand{\phmod}[1]{\left| #1 \right|}
\newcommand{\pr}[1]{\left( #1 \right)}

\newcommand{\bpg}[1]{\biggl\{ #1 \biggr\}}
\newcommand{\lab}[1]{\label{#1}}

\newcommand{\mat}[1]{\underline{\underline{#1}}}
\newcommand{\oo}[1]{\overline{#1}}
\newcommand{\e}[1]{Eq.~\eqref{#1}}
\newcommand{\es}[2]{Eqs.~\eqref{#1}--\eqref{#2}}
\newcommand{\elab}[2]{Eq.(\ref{#1}#2)}

\newcommand{\fig}[1]{Fig.\ref{#1}}

\newcommand{\evalat}[2]{\left.#1\right|_{#2}}
\renewcommand{\sec}[1]{Section\,\ref{#1}}
\newcommand{\app}[1]{Appendix\,\ref{#1}}
\newcommand{\h}[1]{\hat{#1}}

\newcommand{\wt}[1]{\widetilde{#1}}

\newcommand{\ocite}[1]{Ref.\cite{#1}}
\def\grad{\mbox{\boldmath $\nabla$}}

\newcommand{\p}{\prime}           
\newcommand{\bxi}{\boldsymbol{\xi}}
\def\CS   {\mathbb{C}}
\def\ReS  {\mathbb{R}}

\def\ga         {\alpha}

\def\gc         {\gamma}

\def\gd         {\delta}
\def\gD         {\Delta}
\def\gee        {\epsilon}
\def\gl         {\lambda}

\def\go         {\omega}
\def\gO         {\Omega}
\def\gr         {\rho}

\def\gt         {\theta}

\def\zero	{{\mathbf 0}}

\def\yy		{{\mathbf y}}

\def\vv		{{\mathbf v}}

\def\AA		{{\mathbf A}}

\def\GG		{{\mathbf G}}
\def\RR		{{\mathbf R}}

\def\pp		{{\mathbf p}}
\def\xx		{{\mathbf x}}

\def\rr		{{\mathbf r}}
\def\PP		{{\mathbf P}}

\def\JJ		{{\mathbf J}}

\def\uu		{{\mathbf u}}

\def\kk		{{\mathbf k}}
\def\qq		{{\mathbf q}}

\def\jj		{{\mathbf j}}
\def\ee         {{\mathbf e}}

\newcommand{\di}{\mathrm{d}}
\newcommand{\im}{\mathrm{i}}
\def\bcalll{\mbox{\boldmath $l$}}

\def\callA{\mbox{$\mathcal{A}$}}

\def\callC{\mbox{$\mathcal{C}$}}

\def\callH{\mbox{$\mathcal{H}$}}

\def\callP{\mbox{$\mathcal{P}$}}
\def\callQ{\mbox{$\mathcal{Q}$}}

\def\callS{\mbox{$\mathcal{S}$}}
\def\callT{\mbox{$\mathcal{T}$}}
\def\callU{\mbox{$\mathcal{U}$}}

\def\callZ{\mbox{$\mathcal{Z}$}}

\def\bcallA{\mbox{\boldmath $\mathcal{A}$}}
\def\bcallB{\mbox{\boldmath $\mathcal{B}$}}

\def\bcallE{\mbox{\boldmath $\mathcal{E}$}}

\def\bcallJ{\mbox{\boldmath $\mathcal{J}$}}

\newcommand{\cnrism} {Istituto di Struttura della Materia and Division of Ultrafast Processes in Materials (FLASHit) of the National Research Council, via Salaria Km 29.3, I-00016 Monterotondo Stazione, Italy}

\def\sumq{{\sum_\qq}^\p}
\def\BPref{Bohm--Pines\,\cite{Bohm1953}}
\def\BP{Bohm--Pines\,}
\newcommand{\hwt}[1]{\h{\wt{#1}}}
\def\bm{\mathbf}

\newcommand{\mysec}[1]{{\em #1}}
\newcommand{\lux}{Department of Physics and Materials Science, University of Luxembourg, L-1511 Luxembourg, Luxembourg}

\tikzset{
	current/.style={diamond,fill=white,draw,inner sep=0pt,minimum size=10pt},
	delta/.style={circle,draw,fill=white,inner sep=0pt,minimum size=10pt},
}

\begin{document}

\title{The transverse matter Hamiltonian}
\author{Andrea Marini}
\author{Kai Wu}
\affiliation{\cnrism}
\author{Riccardo Reho}
\affiliation{\lux}

\begin{abstract}
Enrico Fermi in 1932 used classical Gauss equation to derive the Coulomb density--density interaction from the longitudinal electromagnetic potential.  In this
work  we extend the Fermi procedure to the transverse component of the vector potential.  By using a fully quantum canonical transformation, we replace the
transverse vector potential with a current--current and current--current--density interactions.  The transformed Hamiltonian is, then, projected in the
fermionic space providing a matter--only, totally incoherent and gauge respecting Hamiltonian. The removal of coherence will avoid the breakdown of perturbation
theory predicted by Haag's theorem, and make possible to introduce, in the transformed space, the diagrammatic approach.  After discussing the implications of
the transverse interactions on the current theories based on the Fermi procedure, we conclude by showing how the transverse Hamiltonian provides the quantum
origin of the electromagnetic longitudinal--transverse splitting of phonons and other elemental excitations.
\end{abstract}
\date{\today}
\maketitle

\section{Introduction}
The workhorse of condensed--matter theory is the non--relativistic Hamiltonian in which electrons and nuclei
interact through an instantaneous Coulomb potential,
\eqgl{I.1}{
\h{H}_{Fermi}=\h{T}+\h{H}_{C},\\
\h{H}_{C}=\frac{1}{2}\int \di\xx\di\yy\, \frac{\h{\gr}\pr{\xx}\h{\gr}\pr{\yy}}{|\xx-\yy|},
}
where $\h{T}$ is the kinetic operator and $\h{\gr}\(\xx\)$ is the total charge density of all fermionic
species, so that \e{I.1} treats the electron--electron, electron--nucleus and nucleus--nucleus interactions on
the same footing. For nearly a century this Hamiltonian has underpinned essentially every first--principles
method: Hartree--Fock, density--functional theory\,\cite{R.M.Dreizler1990} and its perturbative
extension\,\cite{Baroni1987,BaroniRMP}, dynamical mean--field theory\,\cite{Georges1996}, and many--body
perturbation theory\,\cite{Hedin1965,Onida2002}. 
E.~Fermi obtained \e{I.1} in 1932\,\cite{Fermi1932} from the non--relativistic quantum electrodynamics\,(QED)
Hamiltonian,
\mll{I.2}{
\h{H}_{QED}= \h{H}_{EM}
+ \int \di\xx\,\h{A}_0\pr{\xx}\h{\rho}\pr{\xx}+\\
+ \sum_s\frac{1}{2m_s}\int \di\xx\,\h{\psi}_s^\dagger\pr{\xx}\left[-\im\bm\nabla-\frac{\callQ_s}{c}\h{\bm A}\pr{\xx}\right]^2\h{\psi}_s\pr{\xx},
}
by imposing the Coulomb gauge
\eql{I.3}{
	\bm\nabla\cdot\h{\bm A}\pr{\xx}=0.
}
In this gauge the scalar potential $\h{A}_0$ carries no conjugate momentum: it is fixed by Gauss' law and can
be eliminated exactly in favor of the instantaneous interaction $\h{H}_C$. This elimination is gauge
respecting\cite{gauge-comment}, 
and it is precisely what guarantees that \e{I.1} satisfies the continuity equation, the Ward
identities\,\cite{Strinati1988,Stefanucci2025,Rostami2021,Boyack2016}, and produces conserving many--body
approximations\cite{ALEXANDERL.FETTER1971}.

This construction, however, keeps only \emph{half} of the electromagnetic field. Fermi's procedure disposes of the \emph{longitudinal} sector carried by
$\h{A}_0$, but the \emph{transverse} vector potential $\h{\bm A}_\perp$ survives in \e{I.2} through the paramagnetic and diamagnetic couplings. It is almost
universally discarded, on the grounds that magnetic and retardation effects are of order $v/c$ and hence negligible\,\cite{Cohen-Tannoudji2019-3}, so that the
Hamiltonian \e{I.1} is in truth a specific gauge choice describing purely longitudinal physics. This omission is not innocuous. It underlies the long--standing
ambiguity between the ``length'' and ``velocity'' forms of the light--matter coupling\,\cite{Lamb1987,10.1063/5.0306772}. Moreover, disregarding the transverse
vector potential throws away the electromagnetic (radiative) counterpart of the Coulomb interaction: the magnetic, current--current interaction between moving
charges first identified classically by Darwin\,\cite{Darwin1920}.  Earlier quantum treatments of
this transverse channel, by Holstein, Norton and Pincus\,\cite{Holstein1973} and, more recently, by B\`anyai\,\cite{Banyai2021}, proceeded diagrammatically and
neglected the diamagnetic term, thereby breaking gauge invariance.  A microscopic, gauge--respecting, matter--only derivation (the transverse analogue of Fermi's
construction), has remained missing. 

\mysec{Fulfillment of Maxwell equations.}
When we consider the contribution of transverse fields two distinct mechanisms are at work. First, the diamagnetic $\h{\bm A}^2\pr{\xx}$ term of \e{I.2}
renormalizes the photon frequency contained in $\h{H}_{EM}$; as we show later, this renormalization is a
{\em quench} of the free photon (a harmonic oscillator whose frequency is suddenly changed). The quench is
solved exactly by a Bogolyubov transformation, whose ground state is a \emph{squeezed} vacuum, a superposition of arbitrarily many (paired)
photons in the free basis. Second, the paramagnetic coupling to the current \emph{displaces} the field, since
the exact ground state $\ket{\Psi}$ of $\h{H}_{QED}$ must satisfy
$\bra{\Psi}\h{\bm A}\pr{\xx}\ket{\Psi}=\bcallA\pr{\xx}$ (the classical vector potential), a \emph{coherent}
state built on infinitely many photons. 

The exact ground state is therefore a displaced squeezed (squeezed
coherent) state, and either mechanism by itself already requires to use an arbitrary large number of free photons.
This infinite population of photons becomes continuous in the thermodynamic limit where Haag's theorem~\cite{Duncan2012-lh,Haag1955OnQuantumFieldTheories}
states that the $\ket{\Psi}$ is orthogonal to the independent particle state. In other words
no single, universal Hilbert space representation can describe both free and interacting fields. This points to the failure of 
the perturbative expansion.
Even in the finite volume case, where the diagrammatic method can be applied, the consequence is that at any finite
order of the perturbative expansion of $\h{H}_{QED}$, macroscopic Maxwell equations are not satisfied.

In the longitudinal case  the Fermi procedure, enforcing the Gauss` law, removes the condition for $\ket{\Psi}$ to be coherent, and thus
not perturbative. This implies that Fermi procedure is the very basic pillar of the entire Many--Body treatment. It is the removal of the
longitudinal quantum fields that makes possible to introduce a converging  perturbative treatment.

If we now move to the transverse fields, a solution to the failure of the perturbative approach as predicted by Haag's theorem~\cite{Duncan2012-lh,Haag1955OnQuantumFieldTheories}, 
is to extend the Fermi procedure to the transverse field in such a way to shift the Hamiltonian to a new ground state, $\ket{\Psi^\p}$ that, by
construction, respects Maxwell equations.

\mysec{In this work}
, by extending the collective--coordinate approach of \BPref\ to the transverse sector, we construct a canonical
transformation that enforce transverse Amp\`ere equation, thus retaining the diamagnetic term. We will see that this will correspond
to a non perturbative transformation, generalization of the Bogolyubov transformation, that will allow us to project the result
onto the fermionic subspace. This yields a gauge--respecting, matter--only Hamiltonian in which
the transverse photon is replaced by a Yukawa--screened current--current interaction supplemented by a
density--assisted, six--body longitudinal--transverse term. Gauge invariance enters not as an outcome but as a
constraint: it dictates the Yukawa (rather than bare) screening through the diamagnetic term, and forces the
additional six--body interaction to appear.

More importantly, this new Hamiltonian is suited to perturbation theory: the non--perturbative,
infinite-photon dressing required by Maxwell's equations is absorbed once and for all into the canonical
transformation, so that the matter-only Hamiltonian is expanded around the physically correct
(Maxwell-satisfying) reference, the very step that fails for $\h{H}_{QED}$. This is itself a result of the
present work: with the dressing built into the transformation, the transverse, current-mediated interaction can
be treated (if needed) by the same many-body and diagrammatic methods, and to arbitrary order, as the Coulomb
interaction.

The transformed Hamiltonian couples the \emph{total} current of all charged particles, placing the
electron-electron, electron-nucleus and nucleus-nucleus magnetic interactions on the same footing as their
Coulomb counterparts. We organize the applications \sec{sec:applications}) by the pair of currents involved:
the transverse-longitudinal splitting of excitons (electron-electron), and of phonons (nucleus-nucleus); and, since
the interaction strength is set by the transverse vector potential propagator (hence by the electromagnetic environment),
a range of environment-controlled settings (cavity superconductivity and excitons, the quantum Hall effect,
Dirac and topological materials, strongly correlated systems). The total-current formulation also provides a
gauge-consistent starting point for current-density and time-dependent current-density functionals beyond the
clamped-nuclei approximation.

The paper is organized as follows. Due to its paramount importance, in \sec{sec:haag} we will use the diamagnetic term to connect, in a simplified way,
the dressing of the photon state to Haag's theorem. 
\sec{sec:fermi} reviews the Fermi procedure and the elimination of the
longitudinal field. The transverse electromagnetic quantized Hamiltonian is introduced in \sec{sec:T_Hamiltonian}.
In \sec{sec:H_JJ} we introduce the canonical transformation\,(\sec{sec:canonical}) which leads to 
the transverse fermionic Hamiltonian\,(\sec{sec:H_final}) where the paramagnetic and diamagnetic 
interactions are replaced with purely fermionic current--current interactions.
\sec{sec:QED_connection} connects the result to the transverse photon propagator: it shows that the exact ground state is a squeezed coherent state
orthogonal to the free vacuum (so that a perturbative treatment becomes legitimate only after the
transformation), and recovers $\h{H}_{matter}$ as the static limit of the photon exchange diagram.
\sec{sec:applications} presents the applications and implications, and \sec{sec:conclusions} concludes.
The paper also contains several appendixes\,(\app{APP_PLR}--\app{APP_comm}) that provide the reader all ingredients to follow every step of the derivations.

\section{Motivation: the diamagnetic interaction and Haag's theorem}
\lab{sec:haag}
Let's consider, for simplicity, only the diamagnetic contribution to \e{I.2}. After Fourier expanding the 
vector potential operator it turns out that the electromagnetic part of $\h{H}_{QED}$ can be written as
\mll{HAAG.1}{
 \h{H}_{QED}\approx \h{H}^{dia}_{QED}=
 \frac{1}{2} {\sum_{\qq}}^\p \left[\h{\bm e}^\dag_{\qq}\cdot\h{\bm e}_{\qq} + \gO_q^2 \h{\bm a}^\dag_{\qq}\cdot\h{\bm a}_{\qq}\right]=\\=
 \h{H}^0_{QED}+\frac{1}{2} {\sum_{\qq}}^\p  \gO_p^2 \h{\bm a}^\dag_{\qq}\cdot\h{\bm a}_{\qq},
}
where
$\h{H}^0_{QED}$ is the bare photon Hamiltonian and the remind last is the mean--field diamagnetic correction.
In \e{HAAG.1} $\gO_q^2=\(cq\)^2+\gO_p^2$, with $q$ the photon momentum, $c$ the speed of light and $\gO_p$ the Drude frequency\,(\e{TH.14}).
$\gl$ runs over the two transverse photon modes. \e{HAAG.1} will be  demonstrated in detail in \sec{sec:T_Hamiltonian}.

In \sec{sec:QED_bogolyubov} it is demonstrated that
\eqgl{HAAG.2}{
 \h{H}^{0}_{QED}= {\sum_{\qq}}^\p\sum_\gl cq\(\h{b}^{\dag}_{\qq\gl}\h{b}_{\qq\gl}+\tfrac12\),\\
 \h{H}^{dia}_{QED}= {\sum_{\qq}}^\p\sum_\gl\gO_q\(\h{b}^{\p\dag}_{\qq\gl}\h{b}^\p_{\qq\gl}+\tfrac12\).
}
If we now derive exactly the two ground states $\ket{0}$\,($\ket{0^\p}$) of $\h{H}^{0}_{QED}$\,($\h{H}^{dia}_{QED}$) simple
analytical steps\,(\sec{sec:QED_bogolyubov}) lead to 
\eql{HAAG.3}{
 \lim_{V\rar\infty} \bra{0}\ket{0^\p}=0,
}
In \e{HAAG.3} $V$ is crystal volume and $V\rar\infty$ corresponds to the thermodynamic limit.
Clearly \e{HAAG.3} breaks Gell--Mann\&Low theorem~\cite{ALEXANDERL.FETTER1971} and demonstrates that the renormalized ground state $\ket{0^\p}$  is perpendicular to
the unperturbed Fock space and, thus, cannot be reached by using perturbation theory.

This is a simple manifestation of Haag's theorem~\cite{Duncan2012-lh,Haag1955OnQuantumFieldTheories} that has been demonstrated in more general cases, including the fully relativistic
form of $\h{H}_{QED}$.

In this work one of the results will be to embody the renormalization $\(cq\)\Rar \gO_q$ in a non perturbative canonical transformation that will,
automatically, move $\ket{0}$ to $\ket{0^\p}$.

\section{The Fermi procedure and the longitudinal Coulomb interaction}
\lab{sec:fermi}
We consider $S$ different species of classical charged particles interacting with 
electric and magnetic fields. We also assume to be in the Coulomb gauge, \e{I.3}.
We have $N_s$ particles in each species and each particle of kind 
$s=1,\dots,S$ has charge $\callQ_s$ and mass $m_s$. We impose that the system is neutral, $\sum_s N_s\callQ_s=0$.

The particle positions and canonical momenta are denoted by $\xx_{is}$ and $\pp_{is}$, respectively, with
$i=1,\dots,N_s$. All spatial integrals in this section are taken over the full periodic crystal volume $V$.

We split the electromagnetic field into a longitudinal (curl--free) part, rigidly tied to the
instantaneous charge distribution, and a transverse (divergence--free) part, which carries the independent
radiation degrees of freedom. 

By applying the Helmholtz decomposition~\cite{Jackson1998} and using the classical version of \e{I.3} we get
$\bcallA\(\xx,t\)=\bcallA_\perp\(\xx,t\)$ and
\mll{MC.2}{
\bcallE\(\xx,t\) = -\bm\nabla\callA_0\(\xx,t\) -\frac{1}{c} \partial_t \bcallA_\perp\(\xx,t\)=\\= \bcallE_\parallel\(\xx,t\) + \bcallE_\perp\(\xx,t\).
}
The classical Hamiltonian of the system\footnote{In this paper we use electromagnetic Gaussian units}~\cite{reichardt2019theory} is
\begin{widetext}
\eql{MC.H}{
\callH\(t\) = \callU_{EM}\(t\)+ \int_c \di\xx\, \callA_0\(\xx,t\)\rho\(\xx,t\)+
\frac{1}{4\pi} \int_c \di\xx\, \bcallE_\parallel\(\xx,t\)\cdot\bm\nabla \callA_0\(\xx,t\)
+ \sum_{s=1}^{S} \sum_{i=1}^{N_s} \frac{1}{2m_s} \left[\pp_{is}\(t\) - \frac{\callQ_s}{c}\, \bcallA\(\xx_{is},t\)\right]^2.
}
\end{widetext}
$\callU_{EM}$ is the energy stored in the fields; the
second term is the electrostatic coupling of the charges to the scalar potential $\callA_0$; the third is the
longitudinal field energy; and the last, minimal--coupling, term is the particle kinetic energy,
with the vector potential entering through the combination $\pp-\callQ\bcallA/c$. The entire interaction with
the transverse field is therefore hidden inside the last term of \e{MC.H}.
In addition to \e{MC.2}, we have
\eql{MC.3}{
 \bcallB\(\xx,t\) = \bm\nabla\times\bcallA\(\xx\),
}
and the magnetic field is entirely transverse. In \e{MC.H}
\eql{MC.4}{
  \callU_{EM}\(t\)=\frac{1}{8\pi} \int_c \di\xx\, \left[|\bcallE\(\xx,t\)|^2 + |\bcallB\(\xx,t\)|^2\right].
}
The particle density, charge density, and classical paramagnetic current are defined as
\eqgl{MC.5}{
n_s\(\xx,t\) = \sum_{i=1}^{N_s} \delta\(\xx-\xx_{is}\(t\)\),\\
\rho\(\xx,t\) = \sum_{s=1}^{S} \callQ_s\,n_s\(\xx,t\),\\
\bcallJ\(\xx,t\) = \sum_{s=1}^{S} \sum_{i=1}^{N_s} \frac{\callQ_s}{m_s}\, \pp_{is}\(t\)\, \delta\(\xx-\xx_{is}\(t\)\).
}
The entire Hamiltonian \e{MC.H} splits into longitudinal and transverse parts,
\seql{MC.6}{
\eq{
 \callH\(t\)=\callH^\perp\(t\)+\callH^\parallel\(t\),
}
with
\ml{
 \callH^\perp\(t\)=\callU^\perp_{\rm EM}\(t\)+\\+ \sum_{s=1}^{S} \sum_{i=1}^{N_s} \frac{1}{2m_s} \left[\pp_{is}\(t\) - \frac{\callQ_s}{c}\,
 \bcallA\(\xx_{is},t\)\right]^2,
}
and
\ml{
 \callH^\parallel\(t\)=\callU^\parallel_{\rm EM}\(t\)+ \int_c \di\xx\, \callA_0\(\xx,t\)\rho\(\xx,t\)+\\+
 \frac{1}{4\pi} \int_c \di\xx\, \bcallE_\parallel\(\xx,t\)\cdot\bm\nabla \callA_0\(\xx,t\).
}
}
Here $\callH^\parallel$ collects the scalar potential and the longitudinal field energy, while $\callH^\perp$
holds the matter kinetic energy together with the transverse radiation and its coupling to the charges.

The longitudinal block, however, is not an independent part of the dynamics. Since \e{MC.6} contains no time
derivative of $\callA_0$, the scalar potential carries no conjugate momentum: it is a Lagrange multiplier
rather than a dynamical field, and the associated Hamilton equation is not an equation of motion but the Gauss
constraint,
\eql{MC.8}{
 \bm\nabla\cdot\bcallE_\parallel\(\xx,t\) = 4\pi\rho\(\xx,t\),
}
which binds the longitudinal field to the instantaneous charge density. Solving \e{MC.8} for $\callA_0$ and
substituting it back therefore collapses the whole longitudinal block, field energy, coupling and mixed term
alike, into a single instantaneous interaction. Using integration by parts and the Coulomb gauge
$\bm\nabla\cdot\bcallA=0$,
\eql{MC.11}{
 \callH^\parallel\(t\)= \frac{1}{2} \int_c \di\xx \int_c \di\xx^\p\, \rho\(\xx,t\) v_C\(\xx-\xx^\p\) \rho\(\xx^\p,t\),
}
with the bare longitudinal electron--electron interaction
\eql{MC.12}{
 v_C\(\xx-\xx^\p\)= \frac{1}{|\xx-\xx^\p|}.
}
The longitudinal field has thus disappeared as an independent object, surviving only as the Coulomb
interaction between the charges. This is the essence of Fermi's procedure.

What remains in $\callH^\perp$ is the minimal--coupling kinetic term, which we expand as
\mll{MC.13}{
\sum_{s=1}^{S} \sum_{i=1}^{N_s} \frac{1}{2m_s} \left[\pp_{is}\(t\) - \frac{\callQ_s}{c}\, \bcallA\(\xx_{is},t\)\right]^2 =\\= \callT\(t\) + \callH_{jA}\(t\) +
\callH_{A^2}\(t\),
}
into the bare kinetic energy
\eql{MC.14}{
\callT\(t\) = \sum_{s=1}^{S} \sum_{i=1}^{N_s} \frac{\pp_{is}\(t\)^2}{2m_s},
}
and the two field couplings
\eqgl{MC.15}{
\callH_{jA}\(t\) = -\frac{1}{c} \int_c \di\xx\, \bcallJ^\perp\(\xx,t\)\cdot\bcallA\(\xx,t\),\\
\callH_{A^2}\(t\) = \sum_{s=1}^{S} \frac{\callQ_s^2}{2m_s c^2} \int_c \di\xx\, n_s\(\xx,t\) \left[\bcallA\(\xx,t\)\right]^2.
}
The three terms are physically distinct: $\callT$ is the bare kinetic energy; $\callH_{jA}$ is the
paramagnetic coupling through which a moving charge feels the transverse field; and $\callH_{A^2}$ is the
diamagnetic term, quadratic in $\bcallA$. Because $\bcallA=\bcallA_\perp$ in the Coulomb gauge, only the
transverse part of the current couples to the field, as indicated by the projection $\bcallJ^\perp$ in
\e{MC.15}. The diamagnetic term is easily overlooked, and earlier treatments discarded it; yet it is precisely
what protects gauge invariance once the transverse field is eliminated, and retaining it is one of the central
points of this work.

Collecting these terms, the reduced Coulomb--gauge Hamiltonian is
\eql{MC.16}{
\callH\(t\) = U^\perp_{\rm EM}\(t\) + \callT\(t\) + \callH_C\(t\) + \callH_{jA}\(t\) + \callH_{A^2}\(t\),
}
in which the longitudinal sector has been replaced by $\callH_C$ and only the transverse field remains to be
treated.

\section{The transverse Hamiltonian}
\lab{sec:T_Hamiltonian}
The Fermi procedure outlined in \sec{sec:fermi} is based on a classical procedure. The quantization 
\e{MC.16} is performed, indeed, after the longitudinal fields are transformed in the
charge--charge Coulomb interaction.

Here we want to avoid using classical mechanics and transform and remove the transverse vector potential 
by using a fully quantum approach. The quantization of the electromagnetic fields and Hamiltonian
is described in \app{APP_fields}.  
We use the periodic lattice representation defined in
\app{APP_PLR} with the Brillouin Zone\,(BZ) extended to the full reciprocal volume\,(\e{PLR.8}).

The electronic degrees of freedom are quantized by introducing the fermionic field operators
\eqgl{TH.1}{
	\h{\psi}_s\(\xx\) = \frac{1}{\sqrt{V}} \sum_{\kk} e^{\im\kk\cdot\xx} \h{c}_{s\kk},\\
	\h{\psi}_s^\dagger\(\xx\) = \frac{1}{\sqrt{V}} \sum_{\kk} e^{-\im\kk\cdot\xx} \h{c}_{s\kk}^\dagger.
}
\e{TH.1} corresponds to a periodic homogeneous approximation where the electronic band index is absorbed by the 
extended (in the sense of \e{PLR.8}) reciprocal momentum.
From \e{TH.1} it follows that the density and charge--density Fourier components are defined by
\eqgl{TH.2}{
	\h{n}_{s\qq} = \sum_{\kk} \h{c}_{s\kk}^\dagger \h{c}_{s,\kk-\qq},\\
	\h{\rho}_{\qq} = \sum_s \callQ_s \h{n}_{s\qq},\\
	\h{\rho}\(\xx\) = \frac{1}{V} \sum_{\qq} e^{-\im\qq\cdot\xx} \h{\rho}_{\qq}.
}
The paramagnetic current operator is
\eqgl{TH.3}{
	\h{\bm J}_{\qq} = \sum_s \frac{\callQ_s}{m_s}\h{\PP}_{s\qq},\\
        \h{\PP}_{s\qq}= \sum_{\kk} \left(\kk-\frac{\qq}{2} \right) \h{c}_{s\kk}^\dagger \h{c}_{s,\kk-\qq},\\
	\h{\bm J}\(\xx\) = \frac{1}{V} \sum_{\qq} e^{-\im\qq\cdot\xx} \h{\bm J}_{\qq},\\
	\h{\bm J}_{\qq}^{\,\perp} = \mat \callP^\perp\(\qq\) \h{\bm J}_{\qq},
}
with $\(\kk-\qq/2\)$ the average electron-hole momentum. $\mat \callP^\perp\(\qq\)$ is the transverse projector defined in \e{PLR.15}, and the double underline
denotes a Cartesian matrix.

The full Hamiltonian obtained from the Coulomb-gauge Hamiltonian is
\eqgl{TH.9}{
	\h{H} = \h{H}_C + \h{H}^{\perp},\\
	\h{H}^\perp = \h{T}+\h{H}_{\gc}^{\perp} + \h{H}_{jA} + \h{H}_{A^2}.
}
The matter kinetic energy is
\eql{TH.10}{
	\h{T} = \sum_s \sum_{\kk} \frac{k^2}{2m_s} \h{c}_{s\kk}^\dagger \h{c}_{s\kk}.
}
The Coulomb term is
\eql{TH.11}{
	\h{H}_C = \frac{1}{2V} {\sum_{\qq}}^\p \frac{4\pi}{q^2} \h{\rho}_{\qq} \h{\rho}_{-\qq}.
}
The free transverse electromagnetic Hamiltonian is given by \e{QM.11}. The paramagnetic current--field coupling is
\eql{TH.12}{
	\h{H}_{jA} = -\sqrt{\frac{4\pi}{V}} {\sum_{\qq}}^\p \h{\bm J}_{\qq}^{\,\perp} \cdot \h{\bm a}_{\qq}.
}

Finally, the diamagnetic term is
\eql{TH.13}{
	\h{H}_{A^2} = \sum_s \frac{2\pi \callQ_s^2}{m_s V} {\sum_{\qq}}^\p {\sum_{\qq'}}^\p \h{n}_{s(\qq'-\qq)} \h{\bm a}_{-\qq} \cdot \h{\bm a}_{\qq'}.
}

We note that \e{TH.13} can be conveniently rewritten introducing the plasma operator $\h{\gO}_\qq$
\eql{TH.14}
{
 \h{H}_{A^2} =  \frac{1}{2}{\sum_{\qq}}^\p {\sum_{\qq'}}^\p \h{\gO}^2_{\qq'-\qq} \h{\bm a}_{-\qq} \cdot \h{\bm a}_{\qq'},
}
where
\eql{TH.15}
{
 \h{\gO}^2_{\qq} = \sum_s \frac{4\pi \callQ_s^2}{m_s V} \h{n}_{s\qq}.
}

\section{The transverse Coulomb interaction}
\lab{sec:H_JJ}
In 1953 D.Bohm and D.Pines~\cite{Bohm1953} wrote a seminal paper entitled {\em A Collective Description of Electron Interactions: III.  Coulomb Interactions in
a Degenerate Electron Gas} where they formally demonstrate that plasmons can be represented as bosonic modes of a longitudinal quantum electromagnetic field.

In their work \BP introduce the Random Phase Approximation\,(RPA) and demonstrate that there exists a canonical transformation that transforms $\h{H}_{C}$ in a
longitudinal electromagnetic Hamiltonian. Physically they use this equivalence to prove that plasmons are longitudinal electro--magnetic oscillations.

In the following section we will first use the RPA to rewrite the Hamiltonian in a reference, corresponding to a dressed photon propagator,
and a remainder, due to the diamagnetic term. We will then introduce the boundary condition dictated by the Maxwell--Amp\`ere classical condition and, finally, we
will identify the canonical transformation needed to remove all transverse electromagnetic fields.

\subsection{The Random Phase approximation as reference Hamiltonian}
\lab{sec:rpa}
The transverse Hamiltonian is complicated by the diamagnetic term, \e{TH.13}. This term plays an essential role in the 
theory and it is needed to make the final Hamiltonian respecting the gauge constraint.  We start by rewriting \e{TH.15} as
\eqgl{STRPA_M.1}{
 \h{\gO}^2_{\qq} = \gO^2_p\gd_{\qq,\zero}+\gD\h{\gO}^2_{\qq},\\
 \Omega_p^2 = 4\pi\sum_s \frac{\callQ_s^2 \average{\h{n}_{s\zero}}}{V m_s},\\
 \gD\h{\gO}^2_{\qq} = 4\pi\sum_s \frac{\callQ_s^2 \(\h{n}_{s,\qq}-\gd_{\qq,\zero}\average{\h{n}_{s\zero}}\)}{V m_s}.
}
Here 
\eql{STRPA_M.2}{
\average{\h{n}_{s\zero}}\equiv \bra{\Psi} \h{n}_{s\zero} \ket{\Psi},
}
with $\ket{\Psi}$ the exact eigen--ket of $\h{H}$\footnote{\e{CTC.4} can be extended to finite temperature where the single average is replaced by the trace of
the statistical density operator.}. Thanks to \e{STRPA_M.1} it follows
\eql{STRPA_M.6}{
\h{H}_{A^2} =\frac{\Omega_p^2}{2} {\sum_{\qq}}^\p \h{\bm a}_{-\qq} \cdot \h{\bm a}_{\qq}+ 
 \frac{1}{2}{\sum_{\qq}}^\p {\sum_{\qq'}}^\p \gD\h{\gO}^2_{\qq'-\qq} \h{\bm a}_{-\qq} \cdot \h{\bm a}_{\qq'}.
}
The RPA corresponds to assume $\gD\h{\gO}^2_{\qq}=\h{\zero}$ and it is motivated by the fact 
that the  terms in \e{STRPA_M.6} corresponding to different momenta cancel in the
$N\rar\infty$ limit due their rapid oscillations. 

Thanks to \es{STRPA_M.1}{STRPA_M.6} the total transverse photon Hamiltonian, \elab{TH.9}{a},
reduces to
\eqgl{RPA_H}{
 \h{H}^{\perp}= \h{T}+\h{H}_{\gc}^{\prime \perp}+\h{H}_{jA} + \gD\h{H}_{A^2},\\
 \h{H}_{\gc}^{\prime\perp}= \frac{1}{2}{\sum_{\qq}}^\p
 \left[\h{\bm e}_{-\qq}\cdot\h{\bm e}_{\qq}+\Omega_{\qq}^2\,\h{\bm a}_{-\qq}\cdot\h{\bm a}_{\qq}\right],\\
 \gD\h{H}_{A^2}= \frac{1}{2}{\sum_{\qq}}^\p {\sum_{\qq'}}^\p \gD\h{\gO}^2_{\qq'-\qq} \h{\bm a}_{-\qq} \cdot \h{\bm a}_{\qq'}.
}

The independent particle part of \e{RPA_H} defines the diamagnetically dressed transverse-photon frequency
\eql{STRPA_M.8}{
 \Omega_{\qq}^2 = \go_{\qq}^2 + \Omega_p^2.
}

\subsection{The transverse Amp\`ere--Maxwell condition}
Before introducing the canonical transformation we need a key ingredient to interpret physically the results.
Indeed, in the \BP approach an important role is played by the enforcement of the longitudinal Gauss equation. As we are working 
with the transverse part of the electromagnetic fields we need to consider the 
transverse Amp\`ere equation.
We are interested in the equilibrium, static, Hamiltonian. Thus we can ignore the time--dependent electric field contribution
to the classical Amp\`ere equation, which reads
\eql{CTC.2.1}{
\bm\nabla\times\bcallB\(\xx\)-\frac{4\pi}{c}\bcallJ_{kin}^\perp\(\xx\)=0.
}
In \e{CTC.2.1} $\bcallJ_{kin}^\perp$ is the kinematic current~\cite{sakurai1967advanced} which includes the
paramagnetic and diamagnetic contribution. We know, again from Maxwell, that
\eqgl{CTC.2.2}
{
 \grad\cdot\bcallE_\perp\(\xx\)=\bm \zero,\\
 \grad\cross\bcallE_\perp\(\xx\)=\bm \zero.
}
In \e{CTC.2.2} $\bcallE_\perp$ is the \emph{transverse} electric field. We start by consider the quantum mechanics transposition of \e{CTC.2.1}:
\eql{CTC.4}{
 \bra{\Psi}\[ \bm\nabla\times\h{\bm B}\(\xx\)-\frac{4\pi}{c}\h{\bm J}_{kin}^{\perp}\(\xx\)\]\ket{\Psi}=0,
}
and
\eql{CTC.5}{
 \h{\bm J}^{\rm kin}\(\xx\)=\h{\bm J}\(\xx\)-\sum_s\frac{\callQ_s^2}{m_s c}\,\h{\bm A}\(\xx\)\h{n}_s\(\xx\).
}
In \e{CTC.4} we have used the Ehrenfest theorem~\cite{Sakurai2020-cw} to connect the quantum mechanics operators to the classical counterparts.
We can now use the lattice representation (see \app{APP_PLR} and \app{APP_fields}) to write the
$\qq$ component of \e{CTC.4}
\eql{CTC.5.1}{
 \bra{ \Psi}
 \[\h{\bm a}_\qq-\sqrt{\frac{4\pi}{V}} \frac{{\bm J}^\perp_{-\qq}}{\go_{\qq}^2}+
  \sum_{\qq^\p}
  \frac{\h{\gO}^2_{\qq^\p-\qq}}{\go_{\qq}^2}
  \h{\bm a}_{\qq^\p}\]\ket{\Psi}=0.
}
We can now use again the RPA, $\h{\gO}^2_{\qq^\p-\qq} = \gO^2_p\gd_{\qq^\p,\qq}$, to obtain the final form of the Maxwell--Amp\`ere equation:
\seql{MA_conditions}{
\eq{
 \bra{ \Psi}\[ \h{\bm a}_{\qq}-\sqrt{\frac{4\pi}{V}}\frac{\h{\bm J}_{-\qq}^{\,\perp}}{\gO_{\qq}^2}\]\ket{\Psi}=0,
}
while from \e{CTC.2.2} we get the second condition
\eq{
 \bra{ \Psi}
 \bm\ee_\qq
\ket{\Psi}=0,
}}

At this point it is essential to observe that neglecting the diamagnetic term corresponds to replace, in \e{CTC.4} $\h{\bm J}_{kin}$ with
$\h{\bm J}^\perp$ (the paramagnetic contribution) that is not gauge invariant. This would make the Maxwell--Amp\`ere equation not gauge invariant.

We conclude this section by observing that \e{MA_conditions} implies that  $\ket{\Psi}$ is a coherent photon state. Indeed in order to connect
\e{MA_conditions} to classical Maxwell--Amp\`ere equation we must have
\eql{CTC.6}{
 \h{\bm a}_{\qq}\(t\) \ket{\Psi} =\frac{1}{\sqrt{4\pi c^2}} \bcallA_\qq\(t\) \ket{\Psi}.
}
\e{CTC.6} corresponds to the definition of coherent state~\cite{PhysRev.131.2766}. \e{CTC.6} will play an important role when we will
project the full Hamiltonian in the matter--only sub--space.

\subsection{The canonical transformation}
\lab{sec:canonical}
In the previous section, we found that the equilibrium photon field is coherently displaced by the current.
Now, we introduce  a unitary operator $\h{C}$ that shifts the photon back, moving into a frame where the photon sits
in its plain fluctuation vacuum. 

We define
\seql{CTC.8}{
	\eql{CTC.8a}{\h{C}=e^{\im\h{S}},}
	\eql{CTC.8b}{\h{S}=\sqrt{\frac{4\pi}{V}}{\sum_{\qq}}^\p\frac{\h{\bm e}_{\qq}\cdot\h{\bm J}_{\qq}^{\,\perp}}{\Omega_{\qq}^2}.}
}
Since $\h{\bm e}_{-\qq}=\h{\bm e}_{\qq}^{\dagger}$ and
$\h{\bm J}_{-\qq}^{\,\perp}=\left(\h{\bm J}_{\qq}^{\,\perp}\right)^\dagger$,
the generator $\h{S}$ is Hermitian and $\h{C}$ is unitary.

As $\h{C}$ is a functional of $\h{\bm e}_\qq$ it follows that
\eql{CTC.8.0}{
 \h{C}^\dag \h{\bm e}_\qq \h{C}=\h{\bm e}_\qq.
}

By using \e{appA.4} we get
\eql{CTC.8.1}
{
 \hwt{\bm a}_\qq\equiv \h{C}^\dag \h{\bm a}_\qq \h{C}=\h{\bm a}_\qq +\sum_{n=1}^\infty \evalat{\hwt{\bm a}_\qq}{n}.
}
By using the canonical commutation relation, \e{QM.10}, it follows 
\eql{CTC.9}{
 \evalat{\hwt{\bm a}_\qq}{1}=-\im\[\h{S},\h{\bm a}_\qq\]=\sqrt{\frac{4\pi}{V}}\frac{\h{\bm J}_{-\qq}^{\,\perp}}{\Omega_{\qq}^2}.
}
From \e{appA.4} we see that the calculation of $\evalat{\hwt{\bm a}_\qq}{2}$ involves nested, higher order
commutators. As $\evalat{\hwt{\bm a}_\qq}{1}$ is purely fermionic we get 
\eql{CTC.9.1}{
 \evalat{\hwt{\bm a}_\qq}{2}=\frac{1}{2}\[\h{S},\[\h{S},\h{\bm a}_\qq\]\]= \frac{1}{c^4} {\sum_\uu}^\p \h{\mat{\Xi}}_{\qq\,\uu}\cdot \h{\bm e}_{\uu},
}
with $\h{\mat{\Xi}}_{\qq\,\uu}$ a purely fermionic tensorial operator.
By iterating \e{CTC.9.1} it follows that the terms for $n>2$ involve a $n$--order tensorial operator multiplying $(n-1)$ electric field operators.

\begin{widetext}
Let's now use \es{CTC.8.0}{CTC.9.1} to calculate how Maxwell--Amp\`ere condition, \e{MA_conditions} transform:
\eqgl{CTC.9.3}{
0=\bra{ \Psi}\[ \h{\bm a}_{\qq}-\sqrt{\frac{4\pi}{V}}\frac{\h{\bm J}_{-\qq}^{\,\perp}}{\gO_{\qq}^2}\]\ket{\Psi}
= \bra{ \Psi}\h{C}\h{C}^\dag\[ \h{\bm a}_{\qq}-\sqrt{\frac{4\pi}{V}}\frac{\h{\bm J}_{-\qq}^{\,\perp}}{\gO_{\qq}^2}\]\h{C}\h{C}^\dag\ket{\Psi}=
 \bra{\wt{\Psi}}\h{\bm a}_{\qq}+\frac{1}{c^4} {\sum_\uu}^\p \h{\mat{\Xi}}_{\qq\,\uu}\cdot \h{\bm e}_{\uu}\ket{\wt{\Psi}},\\
\bra{\wt{\Psi}}\h{\bm e}_{\qq} \ket{\wt{\Psi}}=0.
}
\end{widetext}
\e{CTC.9.3} reduces to the conditions
\eqgl{CTC.9.4}{
\bra{\wt{\Psi}}\h{\bm a}_{\qq} \ket{\wt{\Psi}}=0,\\
\bra{\wt{\Psi}}\h{\bm e}_{\qq} \ket{\wt{\Psi}}=0.
}
\e{CTC.9.4} contains a crucial and physically sound message: in the transformed Hilbert space the transverse photons must be in a state with zero 
displacement ($\h{\bm a}_{\qq}$) and speed ($\h{\bm e}_{\qq}$). \e{CTC.9.4} also implies that, after the canonical transformation
\eqgl{CTC.10}{
\h{\bm a}_{\qq}=\h{\bm a}_{\qq}-\bra{\wt{\Psi}}\h{\bm a}_{\qq}\ket{\wt{\Psi}}=\gD \h{\bm a}_{\qq},\\
\h{\bm e}_{\qq}=\h{\bm e}_{\qq}-\bra{\wt{\Psi}}\h{\bm e}_{\qq}\ket{\wt{\Psi}}=\gD \h{\bm e}_{\qq}.
}
\e{CTC.10} demonstrates that, physically, the canonical transformation shifts the Hamiltonian in such a way the macroscopic Maxwell equations are
implicitly satisfied. This has a powerful implication as it does not require the transformed ground state to be coherent. Indeed, in connecting 
the canonical transformation to the perturbative expansion of the full Hamiltonian (which corresponds to the non--relativistic quantum--electro dynamics
Hamiltonian) we will realize that it will not possible to recover the canonical transformation as a perturbative series due to the intrinsic non--perturbative
nature of the coherent states.

At the same time the proposed canonical transformation, removing the coherent dynamics, avoids the consequence of Haag's theorem\,(see \sec{sec:haag}
and \sec{sec:QED_bogolyubov}) making possible to perform perturbative treatments in the new,
transformed, Fock space.


\subsection{Effective Hamiltonian}
\lab{sec:H_final}
From \e{CTC.10} it follows that
\eql{T_H.1}{
 \h{C}^\dagger \h{H} \h{C}=\h{H}_{matter}+\gD\h{H}_{matter}\[\gD\h{\bm a}_{\qq},\gD\h{\bm e}_{\qq}\].
}
\e{T_H.1} defines the matter--only Hamiltonian as the canonically transformed Hamiltonian where all
quantum fluctuations of the electromagnetic field are neglected
\eql{T_H.2}{
 \h{H}_{matter}=\evalat{\h{C}^\dagger \h{H} \h{C}}{\gD\h{\bm a}_{\qq}=\gD\h{\bm e}_{\qq}=\h{\bm 0}}.
}
In physical terms \e{MA_conditions} only constrains the field
\emph{expectation values} $\bra{\wt\Psi}\gD\h{\bm a}_{\qq}\ket{\wt\Psi}=\bra{\wt\Psi}\gD\h{\bm e}_{\qq}\ket{\wt\Psi}=0$,
whereas \e{T_H.2} sets the fluctuation \emph{operators} themselves to zero. Passing from the former to the
latter is the defining approximation of the matter--only construction: it discards the
dressing of the photon fields. This is fully consistent with the longitudinal interaction that, within the Fermi procedure
outlined in \sec{sec:fermi}, approximates the charge--charge interaction with the adiabatic limit of the 
Klein--Gordon longitudinal photon field.
$\h{H}_{matter}$, therefore, retains the instantaneous, current--mediated exchange but drops dynamical effects.
This will be demonstrated in \sec{sec:QED_connection}.

In order to calculate $\h{H}_{matter}$ we  need to apply \es{CTC.8.0}{CTC.9.1} to the different terms of $\h{H}$.
We start by calculating the action of $\h{C}$ on the density and kinetic energy operators:  $\hwt{\gr}_{\qq}$ and $\hwt{T}$.

By applying the procedures described in \app{APP_canonical} and \app{APP_comm} it is tedious but simple math to demonstrate that
\eql{T_H.6}{
 \left[\h{\bm J}_{\qq}^{\,\perp}, \h{n}_{s\uu}\right] = \frac{\callQ_s}{m_s} \mat P^\perp\(\qq\)\uu\, \h{n}_{s(\qq+\uu)},
}
and, consequently,
\eql{T_H.7}{
\evalat{\hwt{n}_{s\uu}}{1} =
- \im \sqrt{\frac{4\pi}{V}} \frac{\callQ_s}{m_s} {\sum_{\qq}}^\p
\frac{\gD\h{\bm e}_{\qq}\cdot \mat P^\perp\(\qq\)\uu}{\Omega_{\qq}^2}
\h{n}_{s(\qq+\uu)}
}
In \e{T_H.7} it appears $\gD\h{\bm e}_{\qq}$ and, thus, if we neglect the electromagnetic field quantum fluctuations we get
\eqgl{T_H.8}{
 \[\h{C}^\dagger \h{T} \h{C}\]_{matter}  = \h{T},\\
 \[\h{C}^\dagger \h{H}_C \h{C}\]_{matter}  = \h{H}_C.
}

The same procedure can be applied to the transverse part of the Hamiltonian, as defined in \e{RPA_H}. 
If we apply the canonical transformation to $\h{H}^{\perp \p}_{\gc}$ we get
\eqgl{T_H.9}{
 \evalat{\hwt{H}_{\gc}^{\prime\perp}}{matter}=\frac{2\pi}{V} {\sum_{\qq}}^\p \frac{\h{\bm J}_{\qq}^{\,\perp} \cdot \h{\bm
J}_{-\qq}^{\,\perp}}{\Omega_{\qq}^2},\\
 \evalat{\hwt{H}_{jA}}{matter}=-\frac{4\pi}{V} {\sum_{\qq}}^\p \frac{\h{\bm J}_{\qq}^{\,\perp} \cdot \h{\bm
J}_{-\qq}^{\,\perp}}{\Omega_{\qq}^2}.
}
The last term is a density variation--current--current interaction term:
\eql{T_H.11}{
 \evalat{\hwt{H}_{A^2}}{matter}=
 \frac{2\pi}{V}{\sum_{\qq}}^\p {\sum_{\qq^\p}}^\p \gD\h{\gO}^2_{\qq^\p-\qq} \frac{\h{\bm J}^\perp_{-\qq} \cdot \h{\bm J}^\perp_{\qq\p}}{\gO^2_\qq\gO^2_{\qq^\p}}.
}
Collecting all terms, the final effective Hamiltonian is
\mll{T_H.12}{
 \h{H}_{matter} = \h{T} + \h{H}_C - \frac{2\pi}{V} {\sum_{\qq}}^\p \frac{\h{\bm J}_{\qq}^{\perp} \cdot \h{\bm J}_{-\qq}^{\,\perp}}{\Omega_{\qq}^2}+\\+
\frac{2\pi}{V}{\sum_{\qq}}^\p {\sum_{\qq^\p}}^\p \gD\h{\gO}^2_{\qq^\p-\qq} \frac{\h{\bm J}^\perp_{-\qq} \cdot \h{\bm J}^\perp_{\qq\p}}{\gO^2_\qq\gO^2_{\qq^\p}}.
}
The third term of the r.h.s. of \e{T_H.12} defines the \emph{transverse current--current interaction}
\eql{H_JJ}{
 \h{H}_{JJ}\equiv -\frac{2\pi}{V}{\sum_{\qq}}^\p \frac{\h{\bm J}_{\qq}^{\perp} \cdot \h{\bm J}_{-\qq}^{\,\perp}}{\Omega_{\qq}^2},
}
the leading, current--mediated transverse exchange, while the fourth term is the density--assisted
correction $\h{H}_{JJn}$. We can now
move from the lattice representation back to real--space. By using the rules defined
in \app{APP_PLR} it follows that
\begin{widetext}
\seql{T_H_final}{
\ml{
 \h{H}_{matter} = \h{T} +\frac{1}{2} \int \di \xx_1\xx_2 \h{\rho}\(\xx_1\) \h{\rho}\(\xx_2\)v_{\parallel}\(\xx_1-\xx_2\)
  -\frac{1}{2c^2}  \int \di \xx_1\xx_2 \[\h{\bm J}^\perp\(\xx_1\)\]^\dag\cdot \h{\bm J}^\perp\(\xx_2\)v_{\perp}\(\xx_1-\xx_2\)+\\+
   \frac{1}{2c^4} \sum_s \frac{\callQ_s^2}{m_s}  \int \di \xx_1\xx_2 \xx_3 \( \h{n}_{s}\(\xx_1\)-\average{\h{n}_{s}\(\xx_1\)}\) 
   v_{\perp}\(\xx_1-\xx_2\)v_{\perp}\(\xx_1-\xx_3\)\[\h{\bm J}^\perp\(\xx_2\)\]^\dag\cdot\h{\bm J}^\perp\(\xx_3\),
}
where
\eqg{
 v_{\parallel}\(\xx\)=\frac{1}{|\xx|},\\
 v_{\perp}\(\xx\)=e^{-\gO_p\frac{x}{c}}\frac{1}{|\xx|}.
}}
\end{widetext}

\begin{figure}
	\centering
	\includegraphics[width=\columnwidth]{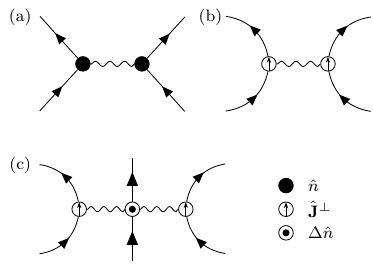}
	\caption{\protect\justifying\footnotesize{
  Diagrammatic representation of the different interaction terms appearing in \e{T_H_final}: (a) the Coulomb interaction, 
 (b) the transverse current--current interaction, and (c) the density-dependent
  current--current interaction.}}
	\label{fig:interaction-diagrams}
\end{figure}

\e{T_H_final} is the central result of our work and collects, in real space, the three matter--only
interaction channels depicted in Fig.~\ref{fig:interaction-diagrams}:
\begin{itemize}
  \item[(a)] the instantaneous longitudinal Coulomb interaction $v_\parallel=1/|\xx|$. This
    is the standard electron--electron, electron--nucleus and nucleus--nucleus interaction; it is the only term 
    kept in, essentially, \e{I.1} and in all ground--state electronic structure.
  \item[(b)] the transverse current--current interaction
    $\propto c^{-2}\,\h{\bm J}^\perp\!\cdot\!\h{\bm J}^\perp\,v_\perp$, the leading magnetic and retardation
    correction. It becomes relevant whenever transverse currents are appreciable: orbital magnetism and
    current--carrying states, the magnetic (Amp\`ere / Darwin--Breit) interaction between moving charges.
  \item[(c)] the density--assisted current--current interaction
    $\propto c^{-4}\,(\h{n}_s-\average{\h{n}_s})\,v_\perp v_\perp\,\h{\bm J}^\perp\!\cdot\!\h{\bm J}^\perp$,
proportional to the \emph{density fluctuation} $\h{n}_s-\average{\h{n}_s}$. It is the
    beyond--RPA approximation employed in the diamagnetic term (the $\gD\h{\gO}^2$ part of \e{STRPA_M.1}) and contributes
    only where strong transverse currents and pronounced density inhomogeneity coexist.
\end{itemize}

Two limits make the physical content of \e{T_H_final} explicit and clarify the role of the plasma
gap $\gO_p$:
\begin{itemize}
  \item \emph{$\gO_p\to 0$ (diamagnetic term switched off).} Then $\gO_{\qq}\to\go_{\qq}=cq$ and
    $\h{H}_{JJ}\propto-\h{\bm J}^\perp\!\cdot\!\h{\bm J}^\perp/(c^2q^2)$, which diverges as
    $\qq\to\zero$. This infrared divergence is the un--screened, long--ranged Biot--Savart/Amp\`ere
    interaction; its appearance signals that the diamagnetic term is indispensable for gauge invariance and
    for the transverse Ward identity.
  \item \emph{$cq\ll\gO_p$ (wavelengths longer than the London length $\gl_L=c/\gO_p$).} Then
    $\gO_{\qq}\simeq\gO_p$, the kernel becomes momentum--independent, and in real space it turns into a constant.
    This imples that 
\eql{T_H.13}{
\evalat{\h{H}_{JJ}}{cq\ll\gO_p}=-\frac{2\pi}{\gO_p^2}\int\di\xx \[\h{\bm J}^\perp\(\xx\)\]^\dag\cdot \h{\bm J}^\perp\(\xx\).
}
\end{itemize}

\mysec{Strength estimation. }
We conclude this section discussing the strength of $\h{H}_{JJn}$ and $\h{H}_{JJ}$. From their definition it may appear that, respectively, 
the $\frac{1}{c^4}$ and $\frac{1}{c^2}$ make the transverse interaction terms relativistically tiny. The appearance of these prefactor is, however, and artifact of the canonical 
transformation and must be used with care.

First of all it is well known that the strength of the $\h{H}_{jA}$ is of the same order of the $\h{\rr}\cdot\bcallE_{ext}\(t\)$ interaction, commonly used to
describe the coupling with laser fields. The two interactions are connected by the G\"oppert--Mayer gauge transformation plus the dipole
approximation~\cite{Cohen-Tannoudji2019-3}.

The G\"oppert--Mayer gauge transformation, thus, introduces a $\frac{1}{c}$ pre--factor without reducing the strength of the interaction.

The present canonical transformation behaves in the same way. If we look at \e{T_H.9} we see that, for example, $\h{H}_{JJ}$ is of the same order of magnitude
of free photons energy, which is not relativistically small. This implies that what dictates the effective importance of the current--current interaction is
the relative strength of $\frac{1}{c}\average{\h{\JJ}^\perp\(\xx\)}$ compared to $\average{\gD\h{n}\(\xx\)}$.

As, from \e{CTC.2.1}, $\frac{1}{c}\average{\h{\JJ}^\perp\(\xx\)}\sim \bm\nabla\times\bcallB\(\xx\)$ we get that the $\h{H}_{JJ}$ and $\h{H}_{JJn}$
can turn particularly important and not negligible in the presence of  strong magnetic fields.

\section{Connection with Quantum Electrodynamics}
\lab{sec:QED_connection}
In the introduction, \sec{sec:haag}, we mentioned that it is not possible to apply straight away perturbation theory on the original Hamiltonian,
\e{RPA_H}, and that the canonical transformation, shifting the dynamics in such a way to embody 
the fulfillment of Amp\`ere equation restores the perturbative nature of the interaction.

In this section we will mathematically see why the perturbative treatment of QED does not have, {\em de facto} access to the exact interacting ground state. 
The transformed Hamiltonian, instead, will have potential access to the exact ground state thanks to
the macroscopic conditions, \e{CTC.10}, imposed by the canonical transformation. 

We will also see that a perturbative treatment will give access to scattering rates and energy renormalization.
This will make possible to realize that 
the physics described by $\h{H}_{matter}$ corresponds to a non--perturbative, adiabatic limit of QED.
Throughout this section, we work with the retarded components of the Green's 
functions on the Keldysh contour~\cite{Stefanucci2025}.

We will start, in \sec{sec:QED_bogolyubov}, discussing the paradigmatic case of the diamagnetic renormalization of
the photon energy. We will then move in \sec{sec:QED_static_limit} to use the lowest order perturbative orders to recover the
terms of the matter--only Hamiltonian \e{T_H_final} as static contractions of the field operators. 
Finally in \sec{sec:QED_PT} we will discuss how
to use perturbation theory to introduce fluctuation corrections to $\h{H}_{matter}$.

\subsection{The diamagnetic mean--field correction}
\subsubsection{The Bogolyubov solution}
\label{sec:QED_bogolyubov}
In \sec{sec:rpa} we embodied the constant plasma diamagnetic frequency in the photon energy, \e{STRPA_M.8}.
A sudden change of an oscillator frequency is a {\em quantum quench}~\cite{Tong_TopicsQuantumMechanics}: it
cannot be treated perturbatively, because the renormalized oscillator has a new ground state containing an
infinite number of un--shifted bosonic excitations.

The proof is simple and the message is so important for the present work that we present the full derivation below.

Renormalizing $\go_q\to\gO_q$ in \e{QM.5}, the free transverse Hamiltonian reads, in terms of dressed
operators,
\eql{SQ.1}{
 \h{H}_{\gc}^{\prime \perp}={\sum_{\qq}}^\p \sum_{\gl} \gO_q \(\h{b}^{\p\dagger}_{\qq\gl} \h{b}^\p_{\qq\gl} + \tfrac{1}{2}\),
}
where the dressed operators are obtained from the bare ones by the Bogolyubov
transformation~\cite{Tong_TopicsQuantumMechanics,tinkham2004introduction}
\eql{SQ.2}{
 \h{b}^\p_{\qq\gl}=u_q\,\h{b}_{\qq\gl}+v_q\,\h{b}^\dag_{\qq\gl},
}
with real coefficients 
\eq{
    u_q=\cosh r_q=\tfrac12\(\sqrt{\gO_q/\go_q}+\sqrt{\go_q/\gO_q}\),
}
\eq{v_q=\sinh r_q=\tfrac12\(\sqrt{\gO_q/\go_q}-\sqrt{\go_q/\gO_q}\),} 
and squeezing parameter
$r_q=\tfrac12\ln\(\gO_q/\go_q\)$. The renormalized ground state is the product of single--mode
squeezed vacua
\eql{SQ.3}{
 \ket{0^\p}=\prod_{\qq\gl}\ga_q\, e^{-\frac{1}{2}\tanh\(r_q\)\[\h{b}_{\qq\gl}^\dag\]^2}\ket{0},
}
each a superposition of even (paired) photon--number states $\ket{2n_{\qq\gl}}$ of the undressed field, with
\eql{SQ.4}{
 \ga_q^2=\frac{1}{\cosh r_q}=\frac{2\sqrt{\go_q \gO_q}}{\go_q+\gO_q}.
}

The overlap is the product of the per--mode factors $\ga_q<1$,
\eql{SQ.5}{
 \bra{0}\ket{0^\p}=\prod_{\qq\gl} \ga_q=\exp\[{\sum_\qq}^\p\sum_{\gl} \ln\ga_q\],
}
whose logarithm, in the thermodynamic limit and summing the two transverse polarizations, becomes the extensive
integral
\eql{SQ.6}{
 {\sum_\qq}^\p\sum_\gl \ln\ga_q = \frac{2V}{\(2\pi\)^3}\int_0^{\infty}\! 4\pi q^2\,\di q\,\ln\ga_q .
}
We prove it finite by treating its two ends explicitly, using $\ga_q^2=2\sqrt{\go_q\gO_q}/\(\go_q+\gO_q\)$ with
$\go_q=cq$ and $\gO_q=\sqrt{c^2q^2+\gO_p^2}$.

At the ultraviolet end ($cq\gg\gO_p$) set $\epsilon\equiv\gO_p^2/c^2q^2\to0$, so that $\gO_q=cq\sqrt{1+\epsilon}$
and
\eqgl{SQ.7}{
 \ga_q^2=\frac{2\(1+\epsilon\)^{1/4}}{1+\sqrt{1+\epsilon}}=1-\frac{\epsilon^2}{32}+O\(\epsilon^3\),\\
 \ln\ga_q=-\frac{\epsilon^2}{64}+O\(\epsilon^3\)=-\frac{\gO_p^4}{64\,c^4q^4}+O\(q^{-6}\).
}
The integrand $q^2\ln\ga_q\sim -q^{-2}$ is integrable as $q\to\infty$, so the ultraviolet end converges.

At the infrared end ($cq\ll\gO_p$) set $\eta\equiv cq/\gO_p\to0$, so that $\gO_q\to\gO_p$ and $\go_q=\gO_p\eta$,
giving
\eqgl{SQ.8}{
 \ga_q^2=\frac{2\sqrt{\eta}\,\(1+\eta^2\)^{1/4}}{\eta+\sqrt{1+\eta^2}}=2\sqrt{\frac{cq}{\gO_p}}\,\[1+O\(\eta\)\],\\
 \ln\ga_q=\tfrac14\ln\!\(\frac{cq}{\gO_p}\)+\tfrac12\ln2+O\(\eta\)\xrightarrow[q\to0]{}-\infty .
}
The divergence is only logarithmic, and the phase--space factor $q^2$ removes it: near $q=0$,
\mll{SQ.9}{
 \int_0^{\gd} q^2\,\di q\,\ln\ga_q\simeq\frac14\int_0^{\gd}\! q^2\ln\!\(\frac{cq}{\gO_p}\)\di q=\\=\frac{\gd^3}{12}\[\ln\!\(\frac{c\gd}{\gO_p}\)-\frac13\],
}
which is finite. Both ends converging, the integral \e{SQ.6} is a finite, negative number, $-\eta\,V$ with $\eta>0$,
so that the overlap decays exponentially with the system size,
\eql{SQ.10}{
 \bra{0}\ket{0^\p}=e^{-\eta V}\xrightarrow[V\to\infty]{}0 .
}
$\ket{0^\p}$ is thus a \emph{squeezed} vacuum, built from the bare vacuum by creating \emph{pairs} of
undressed photons. 
On a finite lattice the overlap is exponentially small but nonzero:
orthogonality is strictly a thermodynamic--limit statement. This is the orthogonality of inequivalent vacua
familiar from the Van Hove model~\cite{VanHove1952} and Anderson's orthogonality catastrophe~\cite{Anderson1967},
and it implies that the exact ground state of $\h{H}_{QED}$ cannot be reached perturbatively from $\ket0$. As
shown in \sec{sec:QED_static_limit}, the interaction \emph{energy} is nonetheless recovered order by order,
because it is an extensive \emph{sum} of per--mode contributions whereas the overlap is the \emph{exponential} of
that sum.

Two distinct operations act on the photon vacuum. The
\emph{diamagnetic} frequency renormalization $\go_q\to\gO_q$, \e{STRPA_M.8}, \emph{squeezes} the vacuum: 
it changes the \emph{width} of the field distribution, creating photons in pairs. The
\emph{paramagnetic} current coupling $\h{H}_{jA}$ instead \emph{displaces} it: being a linear source, it gives
the field a nonzero mean $\average{\h{\bm a}_\qq}\neq0$ fixed by the Maxwell--Amp\`ere condition
\e{MA_conditions} (the \emph{coherent} photon state $\ket{\Psi}$ of \sec{sec:canonical}). The exact ground state
is therefore a \emph{displaced squeezed} (squeezed coherent) state: squeezing sets its width, the current sets
its center. Both operations are extensive, and each alone gives the exact ground state orthogonal to
$\ket0$; this is why perturbation theory built on $\ket0$ fails, whereas the RPA reference of \sec{sec:rpa}
(which absorbs the squeezing) and the canonical transformation of \sec{sec:canonical} (which absorbs the
displacement) together give controlled access to the interacting ground state.

\subsubsection{The dressed photon propagator}
\label{sec:QED_dyson}
We can now wonder whether, although the evaluation of the ground state requires an infinite number of photons, we can
access the photon energy via standard diagrammatic methods.

The diamagnetically dressed, retarded photon Green's function is defined from the Heisenberg--picture
field operator $\h{\bm a}^\p_\qq(t)$ as
\eql{PP.4}{
  \[D^{\prime 0}_\qq\(t\)\]_{ij} \equiv
  -\im\,\gt(t)\,
  \bra{0^\p}\[\h{a}^\p_{\qq i}\(t\),\h{a}^\p_{-\qq j}\(0\)\]\ket{0^\p}.
}
Using the equations of motion generated by the free transverse RPA Hamiltonian $\h{H}_{\gc}^{\prime\perp}$, \elab{RPA_H}{b},
\eqgl{PP.5}{
  \dot{\h{\bm a}}^\p_\qq = -\h{\bm e}^\p_\qq,\\
  \dot{\h{\bm e}}^\p_\qq = \gO_{\qq}^2\,\h{\bm a}^\p_\qq,
}
one finds
\eql{PP.6}{
\h{\bm a}^\p_\qq(t) = \h{\bm a}^\p_\qq\cos(\gO_{\qq} t) - \h{\bm e}^\p_\qq\sin(\gO_{\qq} t)/\gO_{\qq},
} 
and the commutator follows from \e{QM.10},
\eql{PP.7}{
  \left[\h{a}^\p_{\qq i}(t),\, \h{a}^\p_{-\qq j}(0)\right]
  = -\im\frac{P^\perp_{ij}\(\qq\)}{\gO_{\qq}}\,\sin(\gO_{\qq} t).
}
Its Fourier transform
\mll{PP.8}{
\[D^{\prime 0}_\qq\(\go\)\]_{ij} = \int_{-\infty}^{+\infty} \di t\, e^{\im\go t} \[D^{\p 0}_\qq\]_{ij}\(\qq,t\)=\\=
  \frac{P^\perp_{ij}\(\qq\)}{\(\go+\im\eta\)^2 - \gO_{\qq}^2}.
}
In \e{PP.8} $\eta$ is an infinitesimally small positive number introduced to converge the time integral and allowed
by the Gell--Mann\,\&\,Low theorem~\cite{ALEXANDERL.FETTER1971}.

From \e{PP.8} we see, indeed, that the dressed propagator has its pole at the renormalized photon frequency $\gO_{\qq}$.

\subsection{The static limit and the matter--only Hamiltonian}
\label{sec:QED_static_limit}
From \e{PP.8} it follows that
\eql{PP.16}{
  \[D^{\prime 0}_{\qq}\(0\)\]_{ij}  = -\frac{P^\perp_{ij}\(\qq\)}{\gO_{\qq}^2}.
}
We now consider the diagrammatic expansion of the evolution operator $\callS$~\cite{Stefanucci2025} 
\begin{widetext}
\eql{PP.17}{
 \callS=Tr\bpg{ \callT exp\[-\im\int_{\callC} \di \oo{z} \h{H}_{QED}\(\oo{z}\)\]}\equiv \average{exp\[-\im\int_{\callC} \di \oo{z} \h{H}_{QED}\(\oo{z}\)\]},
}
in powers of $\h{H}_{jA}$,\e{TH.12}. In \e{PP.17} $\callT$ the time--ordering on the Keldysh contour, $\callC$.
At the second order we would get 
\eql{PP.18}{
 \callS_2=\(-\im\)^2\frac{2\pi}{V}{\sum_{\qq,\qq^\p}}^\p \int_{\callC} \di z\di z^\p \average{ \h{\bm J}_{\qq}^{\,\perp}\(z\) \cdot \h{\bm a}_{\qq}\(z\)  
 \h{\bm a}_{-\qq^\p}\(z^\p\)  \cdot \h{\bm J}_{-\qq^\p}^{\,\perp}\(z^\p\)}.
}
Now the connection with the effective interaction $\h{H}_{JJ}$, \e{H_JJ} is obtained by replacing the two $\h{\bm a}$ operators with
their contraction evaluated in the static limit
\mll{PP.19}{
 \callS_2\Rar
 -\im\frac{2\pi}{V}{\sum_{\qq,\qq^\p}}^\p \int_{\callC} \di z \di z^\p \average{ \dots \h{\bm J}_{\qq}^{\,\perp}\(z\)
 \cdot \olrar{D^{\prime 0}_\qq}\(z,z^\p\)\gd_{\qq,\qq^\p}\gd\(z-z^\p\) \cdot \h{\bm J}_{-\qq^\p}^{\,\perp}\(z^\p\)}\approx\\\approx
 \im\frac{2\pi}{V}\sumq \int_{\callC} \di z \average{ \dots \h{\bm J}_{\qq}^{\,\perp}\(z\) 
 \cdot \frac{\mat{P}^\perp\(\qq\)}{\gO_{\qq}^2} \cdot \h{\bm J}_{-\qq}^{\,\perp}\(z\)\dots}=
 -\im \int_{\callC}  \di z \average{\dots \evalat{\hwt{H}_{JJ}\(z\)}{matter} \dots}.
}
\e{PP.19} shows that, indeed, $\h{H}_{JJ}$ appears  with the correct prefactor.

The very same procedure can be repeated at the third order by considering the contractions of $\h{H}_{jA} \h{H}_{A^2} \h{H}_{jA}$
\mll{PP.20}{
 \callS_3=\(-\im\)^3\frac{3}{3!}
 {\sum_{\qq_1\qq^\p_1}}^\p 
 {\sum_{\qq_2\qq_3}}^\p 
 \int_{\callC} \di z_1 \di z_2 \di z_3 
 \average{ 
  \[\sum_s \frac{2\pi \callQ_s^2}{m_s V} \gD\h{n}_{s(\qq_1^\p-\qq_1)}\(z_1\) \h{\bm a}_{-\qq_1}\(z_1\) \cdot \h{\bm a}_{\qq_1^\p}\(z_1\)\]\nl
  \[\sqrt{\frac{4\pi}{V}} \h{\bm J}_{\qq_2}^{\,\perp}\(z_2\) \cdot \h{\bm a}_{\qq_2}\(z_2\) \]
  \[\sqrt{\frac{4\pi}{V}}\h{\bm a}_{-\qq_3}\(z_3\)  \cdot \h{\bm J}_{-\qq_3}^{\,\perp}\(z_3\)\]
 }.
}
The factor $3$ in \e{PP.20} comes from the combinations that, at the third order, produce the same term. This is possible because all fermionic operators in
\e{PP.20} appears in pairs and their position  exchange does not lead to any prefactor.
We now see that the $\h{\bm a}$ operators can be contracted in two, topologically equivalent, ways. This results in 
\eql{PP.21}{
 \callS_3\sim \(-\im\)\sum_s \frac{1}{2}\( \frac{4\pi}{V}\)^2 \frac{\callQ_s^2}{m_s}
 {\sum_{\qq\qq^\p}}^\p
 \int_{\callC} \di z
 \average{
  \frac{\gD\h{n}_{s(\qq^\p-\qq)}\(z\)}{\gO_\qq^2 \gO_{\qq^\p}^2}
  \h{\bm J}_{-\qq}^{\,\perp}\(z\) \cdot \h{\bm J}_{\qq^\p}^{\,\perp}\(z\)}= -\im \int_{\callC}  \di z \average{\dots \evalat{\hwt{H}_{A^2}\(z\)}{matter} \dots}.
}
\end{widetext}
\es{PP.19}{PP.21} demonstrates that the result of the canonical transformation coincides to contract all photon field operators and take
the static limit of the corresponding free, diamagnetically dressed, propagators. The two surviving static
vertices are the transverse current--current interaction $\h{H}_{JJ}$, from the second--order contraction of a
single photon line, and its density--assisted correction $\h{H}_{JJn}$, from the third--order contraction of two
photon lines through a $\Delta\h{n}$ insertion, i.e.\ the interactions (b) and (c) of
Fig.~\ref{fig:interaction-diagrams}.


\subsection{Perturbative expansion around $\h{H}_{matter}$.}
\lab{sec:QED_PT}
We conclude the link with QED remarking another crucial result of our work. Since the canonical
transformation has absorbed exactly this infinite-photon dressing, the matter-only Hamiltonian is a legitimate
reference for perturbation theory. Standard Many--Body and diagrammatic expansions can therefore be built on top
of \e{T_H_final}, to arbitrary order, just as on the Coulomb Hamiltonian, whereas the same expansion applied
directly to $\h{H}_{QED}$ around $\ket0$ is obstructed by the orthogonality above.

At this point we notice a key difference compared to the Fermi procedure. As well known from classical mechanics,
and reminded in \sec{sec:fermi},
Gauss law is not a dynamical Hamiltonian equation but, rather, a constrain on the dynamics. Transverse Amp\`ere equation, instead,
is a dynamical equation and, as a consequence, leads to corrections to the static limit of QED, embodied in $\h{H}_{matter}$.

Mathematically this crucial difference between Amp\`ere and Gauss leads to the second term on the r.h.s. of \e{T_H.1}, $\gD\h{H}_{matter}\[\gD\h{\bm
a}_{\qq},\gD\h{\bm e}_{\qq}\]$. The first non vanishing term of $\gD\h{H}_{matter}$ can be derived by applying \e{CTC.9.1}  to \e{TH.12}. The result is
\eql{QED_PT.1}{
 \gD\h{H}_{matter}=-\sqrt{\frac{4\pi}{V}}\frac{1}{c^4}{\sum}^\p_{\qq\uu} \h{\JJ}^\dag_\qq\cdot
\h{\mat{\Xi}}_{\qq\uu}\cdot\gD\h{\ee}_\uu+\cdots.
}
$\(\cdots\)$ represent the other terms coming from the higher--order canonical transformation of the QED Hamiltonian.

Clearly \e{QED_PT.1} represents an interaction term that involves the transverse photon propagator that, thus, admits a 
perturbative expansion that, as highlighted previously, does not suffer of the convergence problems due to Haag's theorem. It is, then, conceptually possible to
introduce elctro--magnetic quantum fluctuations beyond the static limit of \e{T_H_final}.
\section{Applications and implications}
\label{sec:applications}
\e{T_H_final} opens the path to the inclusion of electro--magnetic transverse effects with applications and implications in many
fields.

In \sec{sec:applications_implications} we will discuss what \e{T_H_final} implies for the existing, purely
longitudinal, first--principles theories: the gauge fixing imposed when the system is coupled to external
classical fields, and the internal transverse current--current interaction that those theories omit.

After these implications we, in \sec{sec:LT_splitting}, will discuss how \e{T_H_final} imposes a new interpretation of the  well--known longitudinal--transverse 
splitting\,(LTS) of excitons and phonons, routinely calculated in material science~\cite{Sohier2017,Echeverry2016}.

\begin{figure*}
\centering
\includegraphics[scale=0.5]{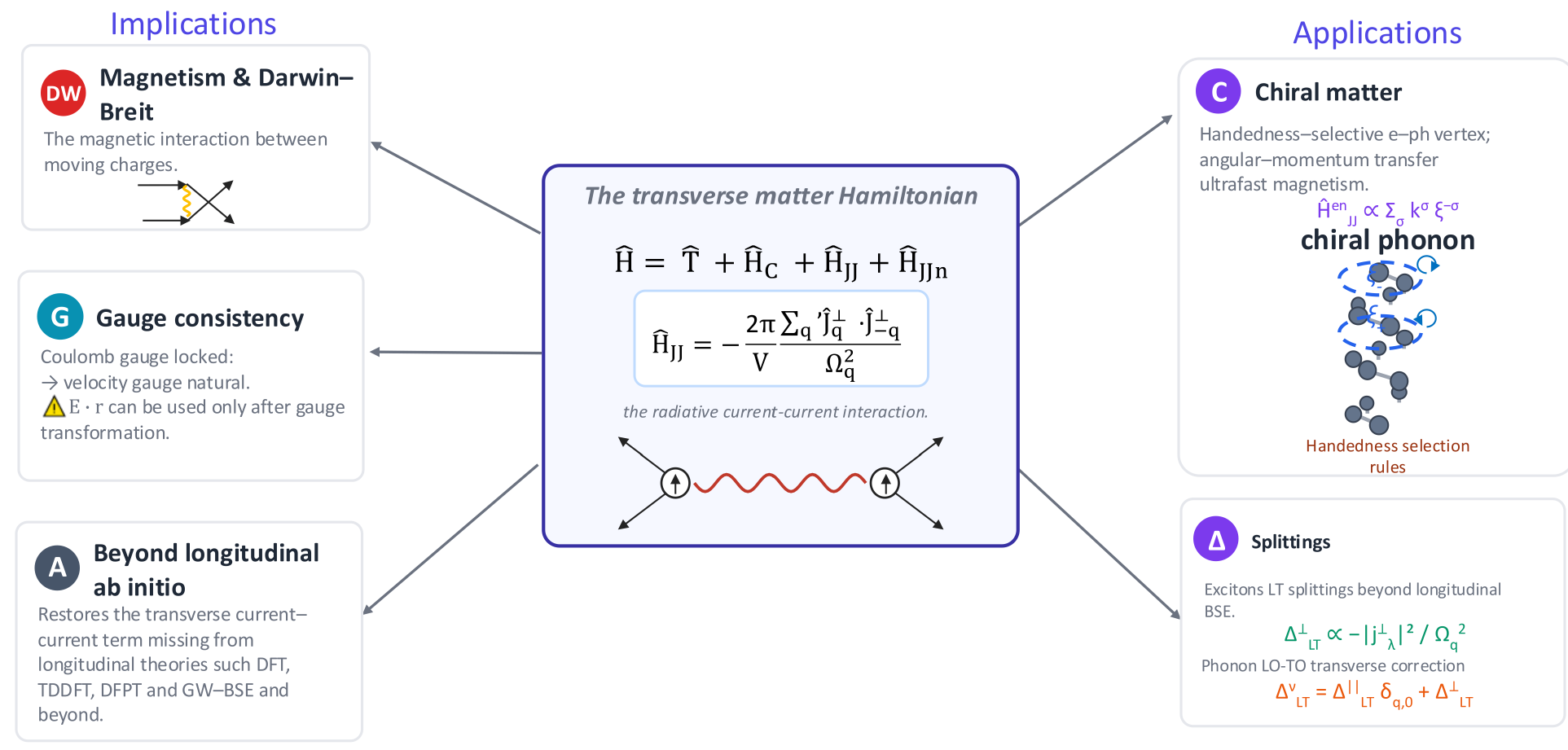}
\caption{\protect\justifying\footnotesize{Visual summary of the implications and applications we envision
as an outcome of the transverse Hamiltonian formalism.}}
\label{fig:vision}
\end{figure*}

Finally, in \sec{sec:chiral_CISS} we discuss the coupling between
electronic currents and chiral lattice motion. We derive the transverse
electron--phonon vertex encoded in $\h{H}_{JJ}$ and show that it is \emph{handedness--selective}, coupling a chiral
phonon only to the electronic current of matching circular sense. On this basis we discuss the implications of our
formalism for \emph{chiral phonons}.
We summarize the implications and applications of our theoretical construction in \fig{fig:vision}.

\subsection{Implications for existing theories}
\label{sec:applications_implications}
\begin{figure*}
\centering
\includegraphics[scale=0.5]{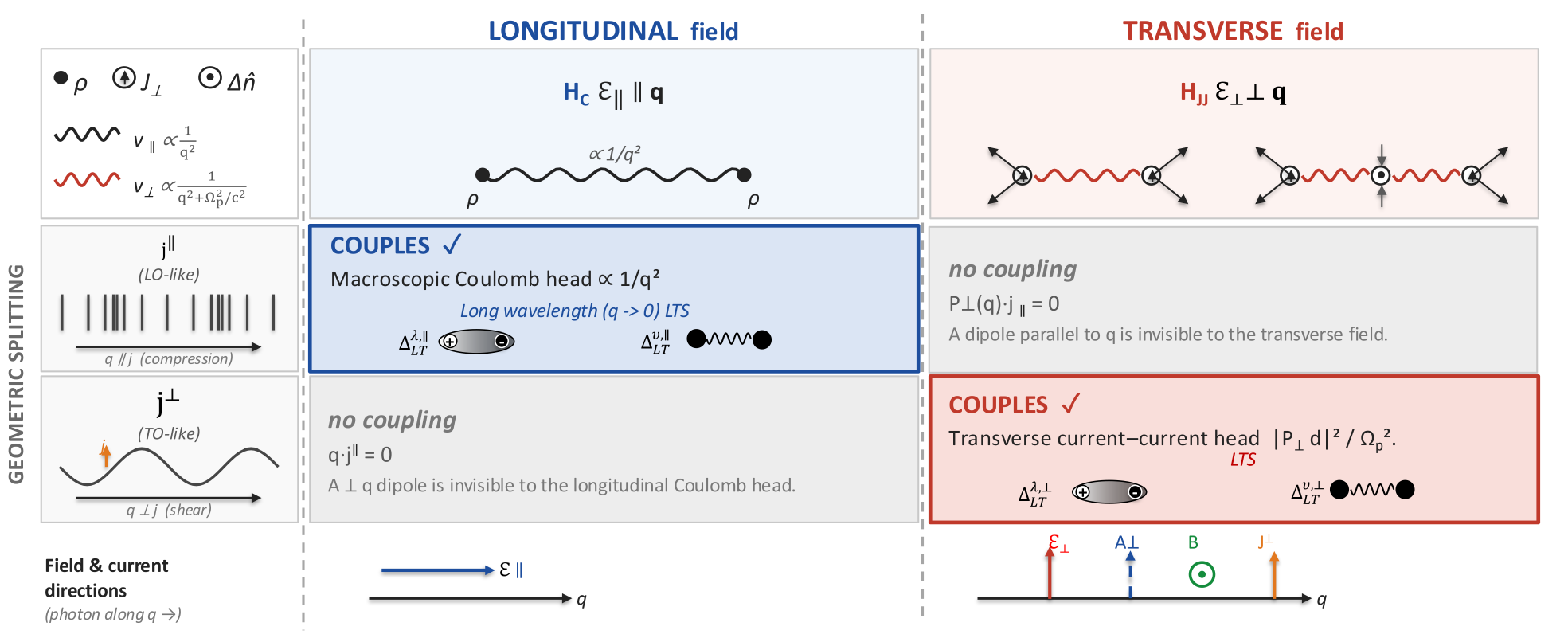}
\caption{\protect\justifying\footnotesize{The two independent ``longitudinal/transverse'' distinctions in play.
\emph{Columns}: the gauge (field) split of the interaction into longitudinal Coulomb ($\h{H}_C$) and
transverse photon ($\h{H}_{JJ}$) sectors. \emph{Rows}: the geometric orientation of the transition current
$\mathbf{j}$ relative to $\qq$. }}
\label{fig:splittings}
\end{figure*}

Nearly every first--principles method of electronic--structure theory is built on the longitudinal Fermi
Hamiltonian \e{I.1}. Density--functional theory in its static~\cite{R.M.Dreizler1990} and
time--dependent~\cite{Marques2006} forms, density--functional perturbation theory for lattice
dynamics~\cite{BaroniRMP}, and the many--body hierarchy~\cite{ALEXANDERL.FETTER1971} that runs through Hedin's
equations~\cite{Strinati1988} to the $GW$ and Bethe--Salpeter formalisms~\cite{Onida2002} are all, by
construction, purely longitudinal. The only partial exception is current--density--functional
theory~\cite{Vignale2006}, which amends \e{I.1} with the {\em external} classical transverse vector potential
$\bcallA_{ext}\(\xx,t\)$, the Runge--Gross theorem~\cite{PhysRevLett.52.997} being extended to the paramagnetic
current~\cite{PhysRevB.70.201102,PhysRevA.38.1149}; even there the {\em internal} transverse quantum field is
absent.

Two implications follow. The first concerns gauge. The Fermi procedure opens with a Helmholtz
decomposition~\cite{Jackson1998} that fixes the Coulomb (radiation) gauge, \e{I.3}, from the outset, and this
fixing is inherited by \e{I.1}; it dictates how the system may be coupled to an external field. Two coupling
forms are in common use~\cite{Lamb1987}: the {\em velocity gauge}, where the internal potential is shifted by the
external one, $\h{\AA}\Rar\h{\AA}+\bcallA_{ext}$, giving $\h{H}_{pA}\equiv\h{H}_{jA}\(\bcallA_{ext}\)$,
\e{TH.12}; and the {\em length gauge}, $\h{H}_{rE}\sim\h{\xx}\cdot\bcallE_{ext}$, reached by a G\"oppert--Mayer
transformation and the dipole approximation~\cite{Cohen-Tannoudji2019-3}. The two agree only for exact
eigenstates; in a truncated basis they differ, which is the origin of the long--standing velocity/length
ambiguity~\cite{Lamb1987} anticipated in the introduction. Since the Coulomb gauge is locked, the velocity gauge
is the natural, gauge--respecting choice, whereas the length gauge is not a relabeling of the Hamiltonian alone:
the G\"oppert--Mayer transformation acts on the field operators and the Hilbert space, and hence on the
self--energies, Green's functions and response kernels derived from them~\cite{Strinati1988}. Consistency then
demands that the transformation must be propagated through the whole many--body apparatus, not applied to \e{I.1} only.
Neglecting this is precisely what makes velocity and length--gauge results disagree in a finite basis.

The second implication is more fundamental. Being longitudinal, none of these theories contains the {\em
internal} transverse current--current interaction $\h{H}_{JJ}$: the radiative counterpart of the Coulomb term is
simply missing. The transverse Hamiltonian \e{T_H_final} restores it, in the same velocity/Coulomb gauge, so
that the internal quantum field and the external classical field are treated on one gauge--consistent footing;
it thus supplies from first principles what current--density--functional theory approximates from the outside.


\subsection{Electro--magnetic longitudinal--transverse splitting of excitons and phonons}
\label{sec:LT_splitting}
There is a rich literature of works dealing with the concept of longitudinal--transverse splitting\,(LTS). 
In the case of electronic excitations Denisov and Makarov~\cite{Denisov1973} work presents a detailed
theory in the case of excitons. Also in the nuclear case the phononic LTS is well known and studied~\cite{Sohier2017}
by using, for example, Density Functional Perturbation Theory\,(DFPT)~\cite{BaroniRMP}.

The aim of this section is to demonstrate that  \e{T_H_final} defines a new, transverse, contribution
to the LTS.


The transverse current--current kernel of \e{T_H_final} produces a macroscopic head that
splits states according to the orientation of their transition current, in both the electronic and the ionic
channel. 
We consider, for simplicity, a state $\ket\gl$ of energy $E_\gl$ and define
\eql{LT.1}{
 \jj^\perp_{\ga \gl}(\qq)=\bra{\ga}\h{\bm J}^\perp_\qq\ket{\gl},
}
By using \e{LT.1} we can calculate the first order correction to $E_\gl$:
\eql{LT.2}{
 \gD E_{\perp}^{\gl} = \la\gl|\h{H}_{JJ}|\gl\ra= -\frac{2\pi}{V}{\sum_{\qq \ga}}^\p \frac{|\jj^\perp_{\ga\gl}(\qq)|^2}{\gO_{\qq}^2}.
}

The same procedure can be applied to the longitudinal interaction where
\eql{LT.3}{
\gD E_{\parallel}^{\gl}= \la\gl|\h{H}_{C}|\gl\ra = \frac{2\pi}{V}{\sum_{\qq\ga}}^\p \frac{|n_{\ga\gl}(\qq)|^2}{\qq^2},
}
and
\eql{LT.4}
{
 n_{\ga\gl}\(\qq\) \equiv \int\di\xx \la \ga|\h{n}_\qq\(\xx\)|\gl\ra.
}
Within the linear--response approximation $\ket{\ga}=\ket{0}$, and \e{LT.4} reduces, in the electronic case, to the Hartree kernel of the Bethe--Salpeter
equation~\cite{Onida2001}.

\e{LT.1} provides a formal and clean way to define a transverse state:
\eql{LT.4.2}{
 \ket{\gl}=
 \begin{cases}
   \ket{\gl^\parallel} & \jj^\perp_{\ga\gl}(\qq)=\zero\qquad\forall \(\qq,\ga\)\\
   \ket{\gl^\perp} & \jj^\perp_{\ga\gl}(\qq)\neq\zero\qquad\forall \(\qq,\ga\)
 \end{cases}.
}
Thanks to \e{LT.4.2} we can collect \es{LT.2}{LT.3} into
\eql{LT.4.1}{
 \gD E_{\gl}=
 \begin{cases}
   \gD E^\parallel_{\gl} & \ket{\gl}=\ket{\gl^\parallel},\\
   \gD E^\perp_{\gl} & \ket{\gl}=\ket{\gl^\perp}
 \end{cases}.
}
It is then clear that we can formally define the LTS as the energy difference between a purely transverse and purely longitudinal state.

The naming longitudinal--transverse in DFPT follows from the $\qq\rar\zero$  of $\gD E^\parallel_{\gl}\(\qq\)$, where
\eql{LT.5}
{
\lim_{\qq\rar\zero} n_\gl(\qq) \sim \qq_\gee\cdot \la 0|\h{\xx}|\gl\ra.
}
In \e{LT.5} we have used $\qq_\gee$ to represent a small $\qq$. From \es{LT.3}{LT.5} we get that
\eql{LT.6}{
 \evalat{\gD E_{\gl}^{\parallel}}{\qq\rar\zero} = \frac{2\pi}{V}\frac{| \qq_\gee\cdot\xx_\gl |^2}{\qq_\gee^2}.
}
In \e{LT.6} $\xx_\gl=\la 0|\h{\xx}|\gl\ra$.
\e{LT.6} makes evident that, as expected, the longitudinal interaction is non zero only when $\xx_\gl\cdot\qq_s\neq 0$. If
$\xx_\gl$ is perpendicular to $\qq_s$ and, thus, transverse, the corresponding $\gD E^{\parallel}_\gl\(\qq\rar\zero\)=0$,
while $\gD E^{\perp}_\gl\(\qq\rar\zero\)\neq 0$ and well defined.

If we compare \e{LT.2} with \e{LT.3} we see some remarkable differences. First of all $\gD E_{\gl}^{\perp}\(\qq\)<0$ while
$\gD E_{\gl}^{\parallel}\(\qq\)>0$.  Second aspect is that $ \gD E^\perp_{\gl}\(\qq\)$ is well behaved when $\qq\rar\zero$. This means,
among other things, that its contribution to the lattice dynamical matrix is analytical and does not require the introduction of
Born effective charges~\cite{BaroniRMP}.

\begin{figure}
\centering
\includegraphics[width=\columnwidth]{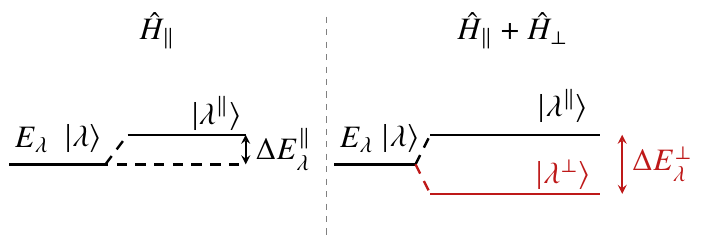}
\caption{\protect\justifying\footnotesize{The geometrical and transverse LT splitting of the exciton energy level. Top, an exciton state eigenvector of a purely longitudinal $\h{H}_{\parallel}$ theory
can split only in the presence of degeneracy, due to $\Delta^{\lambda,\parallel}_{LT} (\qq)$. Bottom, the transverse Hamiltonian $\h{H}_{matter}$ of \e{T_H_final} introduces a new splitting $\Delta^{\lambda,\perp}_{LT} (\qq)$ that is present even in the absence of degeneracy.}}
\label{fig:LTS}
\end{figure}

\es{LT.2}{LT.3} are graphically represented in  \fig{fig:splittings}, while the removal of degenerations caused
by \e{LT.4.1} is represented in \fig{fig:LTS}.

\subsection{Chiral phonons and chirality--induced spin selectivity}
\label{sec:chiral_CISS}

We now turn to the coupling between electronic currents and chiral lattice motion, deriving the explicit
electron--nucleus channel of the transverse interaction $\h{H}_{JJ}$ in the semiclassical, harmonic limit of the
nuclei.  For simplicity we restrict to the case of one species only of atoms with carge $\callZ$.

We start by splitting
transverse current of \e{H_JJ} into its electronic and nuclear parts,
\eql{ate.0}{
 \h{\bm J}^{\perp}_{\qq}\approx\h{\bm J}^{\perp,e}_{\qq}+\bcallJ^{\perp,n}_{\qq}.
}
$\bcallJ^{\perp,n}_{\qq}$ is the classical atomic current at finite momentum that we will define shortly.
We can now use \e{ate.0} to derive the 
electron--nucleus interaction. There will be two terms whith the position of $\h{\JJ}$ and $\bcallJ$ inverted.
These lead to a factor two and
\eql{ate.1}{
 \h{H}_{JJ}^{en}=-\frac{4\pi}{V}{\sum_{\qq}}^\p
 \frac{\oo{\bcallJ}^{\perp,n}_{-\qq}\cdot\h{\bm J}^{\perp,e}_{\qq}}{\gO_{\qq}^2}.
}
The electronic transverse current follows from the paramagnetic current operator \elab{TH.3}{b},
\eql{ate.2}{
 \h{\bm J}^{\perp,e}_{\qq}=\mat\callP^{\perp}\(\qq\) \[\sum_{\kk} \kk\h{c}^{\dagger}_{\kk}\h{c}_{\kk-\qq}\],
}
while for the nuclei we take point ions located at $\h{\RR}_a=\RR_a^0+\h{\uu}_a$, governed by the
phonon Hamiltonian
\eqgl{ate.3}{
 \h{H}_{\rm ph}=\sum_{\qq\gl}\go_{\qq\gl}\(\h{n}_{\qq\gl}+\tfrac12\),\\
 \h{n}_{\qq\gl}=\h{b}^{\dagger}_{\qq\gl}\h{b}_{\qq\gl},
}
with branch index $\gl$ and frequency $\go_{\qq\gl}$. Expanding the displacement in normal modes,
\eqgl{ate.4}{
 \h{\uu}_a=-\frac{\im}{\sqrt{NM}}\sum_{\qq\gl}\bxi_{\qq\gl}\,\h{u}_{\qq\gl}\,e^{\im\qq\cdot\RR_a^0},\\
 \h{u}_{\qq\gl}=\frac{1}{\sqrt2}\(\h{b}_{\qq\gl}+\h{b}^{\dagger}_{-\qq\gl}\),
}
with $N$ the number of cells, $M$ the ionic mass and $\bxi_{-\qq\gl}=\bxi^{*}_{\qq\gl}$, the ionic velocity is
\eqgl{ate.5}{
 \dot{\h{\uu}}_a=\im\big[\h{H}_{\rm ph},\h{\uu}_a\big]
 =\frac{1}{\sqrt{NM}}\sum_{\qq\gl}\bxi_{\qq\gl}\,\go_{\qq\gl}\,\h{p}_{\qq\gl}\,e^{\im\qq\cdot\RR_a^0},\\
 \h{p}_{\qq\gl}=\frac{1}{\sqrt2}\(\h{b}_{\qq\gl}-\h{b}^{\dagger}_{-\qq\gl}\),
}
Now we know that
\eql{ate.5.1}{
 \bcallJ^n\(t\)=\callZ \sum_{a}\dot{\RR}_{a}\(t\)\Rar \h{\JJ}^n=\callZ\sum_{a } \dot{\h{\uu}}_{a}.
}
Last step is to use \e{ate.5} to evalute \e{ate.5.1} and obtain
\eql{ate.6}{
 \bm J^{\perp,n}_{\qq}=\callZ\sqrt{\frac{N}{M}}\sum_{\gl} \bxi^\perp_{\qq\gl}\,\go_{\qq\gl}\,\h{p}_{\qq\gl},
}
where
\eql{ate.6.1}{
 \bxi^\perp_{\qq\gl}=\mat\callP^{\perp}\(\qq\) \bxi_{\qq\gl}.
}
Inserting \es{ate.2}{ate.6} into \e{ate.1} gives the transverse electron--phonon interaction
\eql{ate.7}{
 \h{H}_{JJ}^{en}=\frac{4\pi}{V}\callZ\sqrt{\frac{N}{M}}
 {\sum_{\substack{\qq\gl\\\kk}}}^\p\frac{\go_{\qq\gl}}{\gO_{\qq}^2}
 \[\mat\callP^{\perp}\(\qq\)\kk\] \cdot \bxi^\perp_{\qq\gl}
 \h{p}_{\qq\gl}\h{c}^{\dagger}_{\kk}\h{c}_{\kk-\qq}.
}

\e{ate.7} has the structure of a Fr\"ohlich Hamiltonian but differs in three crucial points: (i) both currents
enter through their transverse component, so only the parts orthogonal to $\qq$ couple and the interaction is carried by
the \emph{transverse} phonons; (ii) it is mediated by the screened transverse--photon propagator $1/\gO_{\qq}^2$ rather
than by the longitudinal Coulomb interaction $1/q^2$, and is hence infrared--finite and short--ranged; and (iii) it is
linear in the phonon momentum $\h{p}_{\qq\gl}$ and therefore couples to the ionic velocity, recovering the magnetic
(Amp\`ere) rather than the electrostatic part of the electron--phonon interaction.

To make the chirality explicit, we introduce a right--handed triad $(\ee_1,\ee_2,\frac{\qq}{q})$ and the
circular polarization vectors $\ee_{\pm}=(\ee_1\pm\im\,\ee_2)/\sqrt2$, which span the plane transverse
to $\qq$ and satisfy $\ee_{\sigma}\cdot\ee^{*}_{\sigma'}=\delta_{\sigma\sigma'}$. Decomposing the transverse
projections in this basis, the kernel of \e{ate.7} becomes diagonal in the circular sector,
\eql{ate.8}{
 \[\mat\callP^{\perp}\(\qq\)\kk\]\cdot\bxi^\perp_{\qq\gl}
 =\sum_{\sigma=\pm}k^{\sigma}_{\qq}\,\xi^{-\sigma}_{\qq\gl},
}
with $k^{\sigma}_{\qq}=\oo{\ee}_{\sigma}\cdot\kk$ and
$\xi^{\sigma}_{\qq\gl}=\oo{\ee}_{\sigma}\cdot\bxi^\perp_{\qq\gl}$. For a chiral mode with (nearly) circular polarization
$\bxi^\perp_{\qq\gl}\!\parallel\!\ee_{\pm}$, carrying $\ell_{\qq\gl}=\pm1$ along $\hat{\qq}$, only one term of \e{ate.8}
survives: the vertex then couples the phonon exclusively to the electronic current of the matching circular sense, so
the transverse electron--phonon coupling is \emph{handedness--selective}.

The central implication of \e{ate.7} is thus that electronic currents couple directly to the transverse,
circular component of the lattice motion. This coupling is absent from the conventional longitudinal electron--phonon
Hamiltonian, which couples to the electrostatic displacement through $\qq\cdot\bxi_{\qq\gl}$ and therefore does not
capture the angular--momentum content of chiral motion. Several open questions remain, for instance, a neutral
quasiparticle such as a phonon cannot couple to a magnetic field through the classical Lorentz force, and the coupling
channel derived here may help clarify how such angular--momentum exchange proceeds microscopically.

Importantly, the vertex itself is spin independent: spin selectivity can emerge only when this orbital
angular--momentum transfer is combined with spin--orbit coupling, which converts orbital into spin angular momentum. The
transverse Hamiltonian therefore provides a gauge--consistent microscopic channel for chiral--phonon--mediated spin
physics, and establishes a first--principles starting point for assessing its contribution to CISS.

The transverse interaction $\h{H}_{JJ}$, and in particular its electron--nucleus channel derived above,
provides a microscopic framework for describing lattice chirality and its coupling to electronic motion. Two natural
applications are the physics of \emph{chiral phonons} and \emph{chirality--induced spin selectivity}, which
have attracted considerable experimental and theoretical interest in recent
years~\cite{ZhangNiu2014PRL,ZhangNiu2015PRL,Zhu2018Science,Ishito2023NatPhys,Gohler2011Science,Naaman2019NatRevChem,Evers2022AdvMater,Bloom2024ChemRev,Fransson2020PRB,Fransson2023PRR}.
Our formalism provides an \ai\ derivation of the microscopic interaction underlying the coupling between electronic
currents and chiral lattice motion, thereby establishing a first--principles route to its quantitative evaluation.

A phonon mode is chiral when the atoms undergo circular or elliptical motion, carrying a finite angular momentum
\eql{CP.1}{
 \bcalll_{\qq\gl}=\im\,\oo{\bxi}_{\qq\gl}\times\bxi_{\qq\gl}.
}
This quantity vanishes for linearly polarized modes, for which
$\bxi_{\qq\gl}$ can be chosen real, and takes the values $\pm 1$ for circularly polarized modes.
Chiral phonons have been predicted at
high--symmetry points of hexagonal lattices~\cite{ZhangNiu2015PRL} and observed in monolayer
WSe$_2$~\cite{Zhu2018Science} and in intrinsically chiral crystals such as $\ga$--HgS~\cite{Ishito2023NatPhys}. Their
angular momentum has also been implicated in lattice--mediated magnetization dynamics and the phononic Einstein--de~Haas
effect~\cite{Tauchert2022Nature}; regarding the latter, we envision that our approach can clarify quantitatively whether
the magnetic moment of an atom is dominated by the orbital motion or by the spin contribution.

\section{Conclusions and outlook}
\lab{sec:conclusions}
We have completed the Fermi procedure. Where Fermi eliminated the longitudinal scalar potential in favor of
the instantaneous Coulomb interaction, we have eliminated the transverse vector potential in favor of an
effective matter--only interaction, using a fully quantum canonical transformation.
Retaining the diamagnetic term throughout, we obtained the
gauge--respecting Hamiltonian \e{T_H_final}, in which the transverse photon is replaced by a Yukawa--screened
current--current interaction together with a density--assisted six--body longitudinal--transverse term. Gauge
invariance is the organizing principle: it fixes the Yukawa (rather than bare) screening through the
diamagnetic contribution and requires the six--body term, so that the result satisfies the same conservation
laws that make the Coulomb Hamiltonian a sound starting point for conserving many--body theory.

A methodological result follows. A direct perturbative treatment of $\h{H}_{QED}$ is obstructed because
its exact ground state, a squeezed coherent state carrying infinitely many photons, is orthogonal to the free
vacuum in the thermodynamic limit. The canonical transformation absorbs precisely this dressing, so that the
matter-only Hamiltonian \e{T_H_final} is itself a legitimate reference for perturbation theory: one can, if
needed, build standard many-body and diagrammatic expansions on top of it, to arbitrary order, exactly as for
the Coulomb Hamiltonian.

The construction unifies the longitudinal and transverse sectors of the electromagnetic field in a single
matter--only Hamiltonian, written in terms of the total current of all charged particles. It therefore places
electron--electron, electron--nucleus and nucleus--nucleus magnetic interactions on the same footing as their
Coulomb counterparts.
We illustrated this with two directly comparable predictions, the transverse
excitonic and phononic longitudinal--transverse splittings, and outlined how the environment dependence of
the propagator opens the transverse interaction to cavity, topological and strongly correlated settings.

Three consequences follow for future first--principles work. First, \e{T_H_final} supplies the transverse,
retarded interaction (phonon-- and exciton--polaritons, orbital--current and magnetic responses) that
longitudinal machinery cannot reach, and which requires the transverse photon propagator rather than the
electrostatic head. Second, because the interaction is expressed through the total current, it offers a
natural, gauge--consistent starting point for current--density and time--dependent current--density
functionals beyond the clamped--nuclei, purely electrostatic approximation. Lastly, the possibility of employing
perturbation theory from a the squeezed quenched vacuum opens the door to a systematic, gauge--consistent treatment 
of the transverse interaction in many--body perturbation theory, 
including chiral phonons. Together, these establish the
transverse matter Hamiltonian as the missing electromagnetic counterpart to the Fermi Hamiltonian that has
underpinned condensed--matter theory for nearly a century.

\section{Acknowledgments}
\label{sec:aknowledgments}
A.M. acknowledges the funding received from: MaX {\em MAterials design at the eXascale}, a European Centre of Excellence funded by the European Union’s program
HORIZON-EUROHPCJU-2021-COE-01 (Grant No.  101093374); {\em Nanoscience Foundries and Fine Analysis -- Europe | PILOT} H2020-INFRAIA-03-2020 (Grant No.
101007417); HPC {\em National Center for HPC, Big Data and Quantum Computing – HPC (Centro Nazionale 01 – CN0000013)}, Decreto Direttoriale MUR n. 1031 del
17/06/2022, registrato dalla Corte dei Conti il 14/07/2022 al n.1880.
R.R. acknowledges the funding received by the FNR through project C22/MS/17415967/ExcPhon.

\appendix
\section{The Periodic Lattice representation}
\lab{APP_PLR}
A crucial tool that we will use throughout the paper is the periodic
lattice representation\,(PLR).

\subsection{Basic definitions}
We label the $N$ unit cells of the finite periodic crystal by the Bravais lattice translations
$\{\RR_a^0\}_{a=1,\dots,N}$.
The volume of the simulation crystal, repeated periodically, is $V=N V_0$.
$V_0$ is, instead, the unit cell volume.

From \ocite{Ashcroft1976} we know an important identity
\eql{PLR.0}{
	\frac{1}{N}\sum_a e^{\im \RR^0_a \cdot \vv}=\gd_{\vv \zero}
}
\e{PLR.0} can be used to represent any function $F\(\xx\)$ as
\seql{PLR.1}{
	\eq{F\(\xx\)=\frac{1}{N}\sum_{\qq\in BZ}e^{\im\qq\cdot\xx}F_\qq\(\xx\),}
	with $\qq$ running in the first Brillouin zone\,(BZ) of the lattice, and
	\eq{F_\qq\(\xx+\RR_a^0\)=F_\qq\(\xx\).}
}
\e{PLR.0} implies that
\eql{PLR.2.0}{
	\frac{1}{N}\sum_{a=1}^N e^{\im\(\qq-\qq'\)\cdot\RR_a^0}=\gd_{\qq\qq'},
}
which leads to
\eql{PLR.2}{
	\int_c e^{\im\qq\cdot\xx}d\xx=\sum_{a=1}^N\int_0 e^{\im\qq\cdot\(\xx+\RR_a^0\)}d\xx=N V_0 \gd_{\qq\zero}.
}
\e{PLR.2} allows to  transform a crystal integral $\int_c$ into a unit--cell integral $\int_0$ and can be used to 
rewrite all integrals.

By using \e{PLR.2.0} we can also define the inverse of \elab{PLR.1}{a}:
\eql{PLR.4}{
	F_\qq\(\xx\) = \sum_{a=1}^N e^{-\im\qq\cdot\(\xx+\RR_a^0\)} F\(\xx+\RR_a^0\).
}
If $F\(\xx\)\in\ReS$, it follows that
\eql{PLR.4.1}{
	F_{-\qq}\(\xx\)=\oo{F_\qq\(\xx\)}.
}

By introducing the reciprocal lattice vectors $\{\GG\}$, we can further expand
the periodic function $F_\qq\(\xx\)$ as
\eql{PLR.5}{
	F_\qq\(\xx\) = \frac{1}{V_0} \sum_\GG F_\qq\(\GG\) e^{\im\GG\cdot\xx}.
}
The corresponding Fourier coefficients are
\eql{PLR.5.1}{
	F_\qq\(\GG\) = \int_0 e^{-\im\GG\cdot\xx} F_\qq\(\xx\)d\xx.
}
For real $F\(\xx\)$, the coefficients satisfy
\eql{PLR.5.2}{
	F_{-\qq}\(\GG\) = \oo{F_\qq\(-\GG\)}.
}

From \elab{PLR.1}{a} and \e{PLR.5} it follows that we can rewrite the PLR
by summing simultaneously over $\qq$ and $\GG$:
\eql{PLR.6}{
	F\(\xx\) = \frac{1}{V} \sum_{\qq\in BZ} \sum_\GG e^{\im\(\qq+\GG\)\cdot\xx} F_\qq\(\GG\).
}

The PLR admits a well defined thermodynamic limit.
In fact,
\mll{PLR.7}{
\lim_{N\rar\infty} F\(\xx\)= \lim_{N\rar\infty} \frac{1}{N}\sum_{\qq\in BZ} e^{\im\qq\cdot\xx}F_\qq\(\xx\)=\\
= \frac{V}{N}\int_{BZ}\frac{d\qq}{\(2\pi\)^3}e^{\im\qq\cdot\xx} F_\qq\(\xx\)=\\
= \frac{V}{N V_0}\sum_\GG \int_{BZ}\frac{d\qq}{\(2\pi\)^3} e^{\im\(\qq+\GG\)\cdot\xx}F_\qq\(\GG\)=\\
= \sum_\GG \int_{BZ}\frac{d\qq}{\(2\pi\)^3} e^{\im\(\qq+\GG\)\cdot\xx}F_\qq\(\GG\).
}
It is simple algebra to demonstrate that
\seql{PLR.7.1}{
	\ml{\int_c \di \yy F\(\yy\)G\(\yy-\xx\)=\\
	\frac{1}{N} \sum_{\qq\in BZ}
	\left[\int_0 \di \yy F_\qq\(\yy\)G_{-\qq}\(\yy-\xx\)\right]e^{\im\qq\cdot\xx}.}
	\ml{\int_c \di \yy F\(\xx-\yy\)G\(\yy\)=\\
	\frac{1}{N} \sum_{\qq \in BZ}
	\left[\int_0 \di \yy F_\qq\(\xx-\yy\)G_{\qq}\(\yy\)\right]e^{\im\qq\cdot\xx}.}
}

A perfectly equivalent formulation is obtained by embedding the $\GG$ vector in the momentum definition $\qq\rar \qq+\GG$. Thanks to
this \e{PLR.6} as
\eql{PLR.8}{
 F\(\xx\) = \frac{1}{V} \sum_{\qq} e^{\im \qq\cdot\xx} F_\qq,
}
with now the $\qq$ summation extended to the entire reciprocal lattice.

\subsection{Classical fields}
In the case of the Coulomb interaction, \e{MC.12}, \e{PLR.8} implies
\eql{PLR.9.1}{
  v_C\(\xx-\xx^\p\)= \frac{1}{V} \sumq \frac{4\pi}{q^2} e^{\im\qq\cdot\(\xx-\xx^\p\)}.
}
The $\qq=\zero$ is removed from the summation in \e{PLR.10} from the initial condition that the total charge in the system  is zero:
\eql{PLR.10}{
\int_c \di\xx\, \rho\(\xx\) = 0,
}
which is equivalent to exclude the divergent $\qq=\zero$ component.

We use \e{PLR.8} to expand the transverse classical vector potential $\bcallA$:
\mll{PLR.11}{
\bcallA\(\xx,t\) = \frac{1}{\sqrt V} {\sum_{\qq}}^\p \bcallA_{\qq}\(t\)\, e^{\im\qq\cdot\xx}=\\
= \frac{1}{\sqrt V} {\sum_{\qq}}^\p \sum_{\gl=1,2} \bxi_{\qq\gl}\, \callA_{\qq\gl}\(t\)\, e^{\im\qq\cdot\xx}.
}
The Coulomb-gauge condition implies that each Fourier coefficient is transverse,
\eql{PLR.12}{
\qq\cdot\bcallA_{\qq}\(t\)=0.
}
Since $\bcallA\(\xx,t\)$ is real, \e{PLR.4.1} implies
\eql{PLR.13}{
\bcallA_{-\qq}\(t\) = \overline{\bcallA_{\qq}\(t\)}.
}
For the polarization vectors $\bxi_{\qq\gl}$ we use the circular convention~\cite{Cohen-Tannoudji2019-3} where
$\bxi_{\qq\gl}\in\CS^3$ and
\eqgl{PLR.14}{
\qq\cdot\bxi_{\qq\gl}=0,\\
\bxi_{-\qq\gl} = \oo{\bxi_{\qq\gl}}=\bxi_{\qq-\gl} ,\\
\oo{\bxi_{\qq\gl}}\cdot\bxi_{\qq\gl^\p} = \gd_{\gl\gl^\p},
}
with $\gl=\pm 1$. The corresponding completeness relation implies that 
\eql{PLR.15}{
 P^\perp_{ij}\(\qq\) \equiv \sum_{\gl=\pm 1} \left(\oo{\bxi_{\qq\gl}}\right)_i \left(\bxi_{\qq\gl}\right)_j = \gd_{ij} - \frac{q_iq_j}{q^2}.
}
\e{PLR.15} defines the transverse projector $P^\perp_{ij}\(\qq\)$.
The potentials Fourier coefficients are defined by
\eqgl{PLR.16}{
 \bcallA_{\qq}\(t\) = \sum_{\gl=1,2} \bxi_{\qq\gl}\, \callA_{\qq\gl}\(t\),\\
 \callA_{\qq\gl}\(t\) = \bxi_{\qq\gl}\cdot\bcallA_{\qq}\(t\).
}
With the convention in \e{PLR.14}, the reality condition \e{PLR.13} becomes
\eql{PLR.17}{
 \callA_{-\qq\gl}\(t\) = \overline{\callA_{\qq\gl}\(t\)}.
}
The magnetic field is
\seql{PLR.18}{
\ml{
\bcallB\(\xx,t\) = \bm\nabla\times\bcallA\(\xx,t\)=\\
	= \frac{\im}{\sqrt V} {\sum_{\qq}}^\p \sum_{\gl} \(\qq\times\bxi_{\qq\gl}\)\, \callA_{\qq\gl}\(t\)\, e^{\im\qq\cdot\xx},
}
and the transverse electric field is
\ml{
\bcallE_\perp\(\xx,t\) = -\frac{1}{c}\partial_t\bcallA\(\xx,t\)=\\
 = -\frac{1}{c\sqrt V} {\sum_{\qq}}^\p \sum_{\gl} \bxi_{\qq\gl}\, \frac{\di\callA_{\qq\gl}\(t\)}{\di t}\, e^{\im\qq\cdot\xx}.
}}

\section{Classical electromagnetic fields normal modes and quantization}
\lab{APP_fields}
We review here  the steps to quantize the electromagnetic fields. Even if those concepts can be easily
found in many text--books~\cite{Cohen-Tannoudji2019-3,sakurai1967advanced} they are essential
to understand the present work.

\subsection{Classical normal modes}
$\callU^\perp_{\rm EM}$, the transverse part of \e{MC.4}, can be easily rewritten in terms of the Fourier coefficients $\bcallA_{\qq}\(t\)$. 
By using
\eql{CM_A.7.1}{
	\frac{1}{V} \int_c \di\xx\, e^{\im\(\qq-\qq^\p\)\cdot\xx} = \gd_{\qq\qq^\p},
}
it follows that the magnetic contribution to \e{MC.4} is
\mll{CM_A.8}{
\evalat{U^\perp_{\rm EM}}{B}(t) =
\frac{1}{8\pi} {\sum_{\qq}}^\p \sum_{\gl\gl^\p} \(\qq\times\oo{\bxi_{\qq\gl}}\)\cdot \(\qq\times\bxi_{\qq\gl^\p}\)\\ \oo{\callA_{\qq\gl}\(t\)}\callA_{\qq\gl^\p}\(t\).
}
Using the identity
\eql{CM_A.9}{
(\bm a\times\bm b)\cdot(\bm c\times\bm d) = (\bm a\cdot\bm c)(\bm b\cdot\bm d) - (\bm a\cdot\bm d)(\bm b\cdot\bm c),
}
together with the transversality and orthonormality conditions in \e{PLR.14}, we get
\eql{CM_A.10}{
	\(\qq\times\oo{\bxi_{\qq\gl}}\)\cdot \(\qq\times\bxi_{\qq\gl^\p}\) = q^2\,\gd_{\gl\gl^\p}.
}
Hence
\eql{CM_A.11}{
	\evalat{U^\perp_{\rm EM}}{B}(t) = \frac{1}{8\pi} {\sum_{\qq\gl}}^\p q^2\,|\callA_{\qq\gl}\(t\)|^2.
}
Similarly,
\eql{CM_A.12}{
	\evalat{U^\perp_{\rm EM}}{E}(t) = \frac{1}{8\pi c^2} {\sum_{\qq\gl}}^\p \phmod{\frac{\di{\callA}_{\qq\gl}\(t\)}{\di t}}^2.
}
Therefore,
\eql{CM_A.13}{
\callU^\perp_{\rm EM}\(t\)= \frac{1}{8\pi} {\sum_{\qq\gl}}^\p \[q^2\,|\callA_{\qq\gl}\(t\)|^2\
	+ \frac{1}{c^2}\phmod{\frac{\di{\callA}_{\qq\gl}\(t\)}{\di t}}^2\].
}
In \es{CM_A.11}{CM_A.13} ${\sum_{\qq\gl}}^\p=\sumq\sum_\gl$.

For the free transverse field, the equations of motion imply
\eql{CM_A.14}{
\nabla^2\bcallA\(\xx,t\) = \frac{1}{c^2}\,\partial_t^2\bcallA\(\xx,t\),
}
which gives
\eql{CM_A.15}{
\frac{\di^2\callA_{\qq\gl}\(t\)}{\di t^2}+\go_q^2\callA_{\qq\gl}\(t\)=0,
}
with $\go_q=qc$.

The general solution compatible with the reality condition \e{PLR.13} can be written by
introducing positive-frequency amplitudes $a_{\qq\gl}$,
\eql{CM_A.16}{
	\callA_{\qq\gl}\(t\) = a_{\qq\gl}\,e^{-\im\go_q t} + \overline{a_{-\qq\gl}}\, e^{\im\go_q t}.
}
The reality condition relates $\callA_{\qq\gl}\(t\)$ and $\callA_{-\qq\gl}\(t\)$, but it does not impose
$a_{-\qq\gl}=\overline{a_{\qq\gl}}$.
The amplitudes $a_{\qq\gl}$ and $a_{-\qq\gl}$ are independent positive-frequency amplitudes
corresponding to the two propagation directions.

Substituting \e{CM_A.16} into \e{CM_A.13}, the oscillating cross terms from the magnetic and
electric contributions cancel and  we obtain
\eql{CM_A.17}{
\callU^\perp_{\rm EM} = \frac{1}{2\pi} {\sum_{\qq}}^\p \sum_{\gl} q^2\,|a_{\qq\gl}|^2.
}

We now introduce real normal-mode variables:
\eqgl{CM_A.18}{
X_{\qq\gl} = \frac{1}{\sqrt{4\pi c^2}} \left(a_{\qq\gl} + \overline{a_{\qq\gl}} \right),\\
P_{\qq\gl} = \im\sqrt{\frac{\go_q^2}{4\pi c^2}} \left(\overline{a_{\qq\gl}} - a_{\qq\gl} \right).
}
It follows that
\eql{CM_A.19}{
 \callU^\perp_{\rm EM} = {\sum_{\qq}}^\p \sum_{\gl} \left(\frac{\go_q^2}{2}\,X_{\qq\gl}^2 + \frac{1}{2}\,P_{\qq\gl}^2 \right).
}
This is the classical normal-mode representation of the classical transverse free field Hamiltonian.

\subsection{Quantum field operators}
\e{CM_A.19} shows that each transverse mode $(\qq,\gl)$ is an independent
harmonic oscillator of frequency $\go_q$.
Canonical quantization promotes $X_{\qq\gl}$ and $P_{\qq\gl}$ to operators satisfying the canonical
commutation relations
\eql{QM.1}{
\left[\h{X}_{\qq\gl}, \h{P}_{\qq^\p\gl^\p}\right] = \im\,\gd_{\qq\qq^\p}\gd_{\gl\gl^\p}.
}
We define the annihilation and creation operators by
\eqgl{QM.2}{
\h{b}_{\qq\gl} = \sqrt{\frac{\go_q}{2}}\, \h{X}_{\qq\gl} + \frac{\im}{\sqrt{2\go_q}}\, \h{P}_{\qq\gl},\\
\h{b}_{\qq\gl}^{\dagger} = \sqrt{\frac{\go_q}{2}}\, \h{X}_{\qq\gl} - \frac{\im}{\sqrt{2\go_q}}\, \h{P}_{\qq\gl}.
}
It follows that
\eqgl{QM.4}{
\left[\h{b}_{\qq\gl}, \h{b}_{\qq^\p\gl^\p}^{\dagger}\right] = \gd_{\qq\qq^\p}\gd_{\gl\gl^\p},\\
\left[\h{b}_{\qq\gl}, \h{b}_{\qq^\p\gl^\p}\right] = \left[\h{b}_{\qq\gl}^{\dagger}, \h{b}_{\qq^\p\gl^\p}^{\dagger}\right] = 0.
}

The free transverse-photon Hamiltonian is therefore
\eql{QM.5}{
\callU^\perp_{\rm EM}\(t\) \Rar \h{H}_\gc^\perp = {\sum_{\qq}}^\p \sum_{\gl} \go_q \left(\h{b}_{\qq\gl}^{\dagger} \h{b}_{\qq\gl} + \frac{1}{2} \right).
}
Similarly the classical coefficients defined in \e{CM_A.16} are now
\eql{QM.6.1}{
\callA_{\qq\gl}\(t\) \Rar \h{A}_{\qq\gl} = \sqrt{\frac{2\pi c^2}{\go_q}} \left(\h{b}_{\qq\gl} + \h{b}^\dagger_{-\qq\gl} \right).
}

Thanks to \e{QM.6.1} we can rewrite the quantum vector potential field in the PLR as
\eqsl{QM.7}{
	\h{\bm A}\(\xx\)
	&= \frac{1}{\sqrt V} {\sum_{\qq}}^\p \sum_{\gl}
	\bxi_{\qq\gl}\, \h{A}_{\qq\gl}\, e^{\im\qq\cdot\xx}\\
	&= \sqrt{\frac{2\pi c^2}{V}}\, {\sum_{\qq}}^\p \sum_{\gl}
	\frac{\bxi_{\qq\gl}}{\sqrt{\go_q}}
	\left(\h{b}_{\qq\gl} + \h{b}^\dagger_{-\qq\gl} \right)e^{\im\qq\cdot\xx}\\
	&= \sqrt{\frac{4\pi c^2}{V}}\, {\sum_{\qq}}^\p
	\h{\bm a}_{\qq}\, e^{\im\qq\cdot\xx}.
}
In order to introduce the quantum transverse electric we transform the classical time derivative, \e{MC.2}, in the corresponding
Hamiltonian commutator
\eqsl{QM.8.0}{
\bcallE_\perp\(\xx,t\)=-\frac{1}{c}\partial_t\bcallA\(\xx,t\)\Rar \h{\bm E}_\perp\(\xx\)=\frac{\im}{c}[\h{\bm A}\(\xx\),\h{H}_{\gc}^{\perp}].
}

Here we introduce the re--scaled Fourier coefficients for convenience:
\seqal{QM.7.1}{
	\h{\bm a}_{\qq} &= \sum_{\gl} \frac{\bxi_{\qq\gl}}{\sqrt{2\go_q}} \left(\h{b}_{\qq\gl} + \h{b}^\dagger_{-\qq\gl} \right),\\
	\h{\bm e}_{\qq} &= \im \sum_{\gl} \sqrt{\frac{\go_q}{2}}\, \bxi_{\qq\gl} \left(\h{b}_{\qq\gl} - \h{b}^\dagger_{-\qq\gl} \right).
}
Using \e{QM.4} and the completeness relation \e{PLR.15}, these re--scaled field variables obey the
canonical commutation relations
\eqgl{QM.10}{
	\left[\h{a}_{\qq i}, \h{a}_{\qq^\p j}\right] = 0,\\
	\left[\h{e}_{\qq i}, \h{e}_{\qq^\p j}\right] = 0,\\
	\left[\h{e}_{\qq i}, \h{a}_{\qq^\p j}\right] = \im\, P^\perp_{ij}\(\qq\)\, \gd_{\qq,-\qq^\p}.
}
By using \e{QM.10} to evaluate the commutator in \e{QM.8.0} we get
\eqsl{QM.8}{
	\h{\bm E}_\perp\(\xx\)
	&= \im \sqrt{\frac{2\pi}{V}}\, {\sum_{\qq}}^\p \sum_{\gl}
	\sqrt{\go_q}\, \bxi_{\qq\gl}
	\left(\h{b}_{\qq\gl} - \h{b}^\dagger_{-\qq\gl} \right)e^{\im\qq\cdot\xx}\\
	&= \sqrt{\frac{4\pi}{V}}\, {\sum_{\qq}}^\p
	\h{\bm e}_{\qq}\, e^{\im\qq\cdot\xx}.
}
The magnetic field operator is
\eqsl{QM.9}{
	\h{\bm B}\(\xx\) &= \bm\nabla\times\h{\bm A}\(\xx\)\\
	&= \im \sqrt{\frac{2\pi c^2}{V}}\, {\sum_{\qq}}^\p \sum_{\gl}
	\frac{\qq\times\bxi_{\qq\gl}}{\sqrt{\go_q}}
	\left(\h{b}_{\qq\gl} + \h{b}^\dagger_{-\qq\gl} \right)e^{\im\qq\cdot\xx}\\
	&= \im \sqrt{\frac{4\pi c^2}{V}}\, {\sum_{\qq}}^\p
	\left(\qq\times\h{\bm a}_{\qq} \right)e^{\im\qq\cdot\xx}.
}
From \elab{PLR.14}{b} it follows
$\h{\bm a}_{-\qq}=\h{\bm a}_{\qq}^{\dagger}$ and
$\h{\bm e}_{-\qq}=\h{\bm e}_{\qq}^{\dagger}$.
Therefore $\h{\bm A}\(\xx\)$, $\h{\bm E}_\perp\(\xx\)$, and $\h{\bm B}\(\xx\)$ are Hermitian.

Finally, the free transverse electromagnetic energy becomes
\eql{QM.11}{
	\h{H}_\gc^\perp= \frac{1}{2} {\sum_{\qq}}^\p \left[\h{\bm e}_{-\qq}\cdot\h{\bm e}_{\qq} + \go_q^2\, \h{\bm a}_{-\qq}\cdot\h{\bm a}_{\qq}\right].
}

\section{General canonical transformation relation}
\label{APP_canonical}
Given a general operator $\h{O}$ let's consider
\eql{appA.1}{
	\h{\wt{O}} \equiv e^{-\im\h{S}} \h{O} e^{\im\h{S}}.
}
We start observing that
\eql{appA.2}{
	\hwt{O}= \h{O}+e^{-\im\h{S}}\[\h{O}, e^{\im\h{S}}\],
}
We now assume that $e^{\im\h{S}}$ can be power expanded to get
\eql{appA.3}{
	\hwt{O}= \sum_{n=0}^\infty \evalat{\hwt{O}}{n},
}
with
\eqgl{appA.4}{
\evalat{\hwt{O}}{0}=\h{O},\\
\evalat{\hwt{O}}{n+1}=\(\frac{-\im}{n+1}\) \[\h{S},\evalat{\hwt{O}}{n}\].
}

\section{An elemental commutator}
\label{APP_comm}
A useful identity used in the paper is obtained by calculating the following elemental commutator
among fermionic operators
\eql{APP_B.1}{
	\[\h{c}^\dag_\kk \h{c}_{\kk-\qq},\h{c}^\dag_\pp \h{c}_{\pp-\uu}\]=\h{c}^\dag_\kk \h{c}_{\pp-\uu}\gd_{\pp,\kk-\qq}-\h{c}^\dag_\pp \h{c}_{\kk-\qq}\gd_{\kk,\pp-\uu}.
}
From \e{APP_B.1} we can calculate several elemental commutators.
The first is
\mll{APP_B.2}{
	\left[\h{J}_{\qq\ga},\h{\gr}_\uu\right]
	=\sum_{\kk\pp}\left(\kk_\ga-\frac{\qq_\ga}{2}\right)
	\left[\h{c}^\dag_\kk\h{c}_{\kk-\qq},\h{c}^\dag_\pp\h{c}_{\pp-\uu}\right]=\\
	= \sum_\kk\left[
	\left(\kk_\ga-\frac{\qq_\ga}{2}\right)\h{c}^\dag_\kk\h{c}_{\kk-\qq-\uu}
	-\nl-\left(\kk_\ga-\uu_\ga-\frac{\qq_\ga}{2}\right)\h{c}^\dag_\kk\h{c}_{\kk-\qq-\uu}
	\right]=u_\ga\h{\gr}_{\qq+\uu}.
}
The second identity is
\eql{APP_B.3}{
	\sum_\pp\[\h{c}^\dag_\kk \h{c}_{\kk},\h{c}^\dag_\pp \h{c}_{\pp-\uu}\]=\h{c}^\dag_\kk \h{c}_{\kk-\uu}-\h{c}^\dag_{\kk+\uu} \h{c}_{\kk}.
}
It also follows that
\eql{APP_B.4}{
	\[\h{n}_\qq,\h{n}_{\uu}\]=\h{0}
}

\bibliography{paper}

\begin{thebibliography}{54}
\bibitem [{gau()}]{gauge-comment}%
  \BibitemOpen
  \href@noop {} {}\bibinfo {note} {A gauge respecting Hamiltonian is not itself
  gauge--invariant but produces gauge--invariant physics. This means that any
  gauge transformation can change the form of the Hamiltonian and of the
  corresponding Green's functions and self--energies. However the basic
  Many--Body equations, like Dyson equation, are left unchanged. This leads to
  gauge independent final results. In this sense we say that the results of
  this work are gauge--respecting.}\BibitemShut {Stop}%
\end{thebibliography}

\end{document}